\documentclass[apj]{emulateapj}
\usepackage{epsfig}
\usepackage{amsmath}
\usepackage{natbib}
\usepackage{graphicx,subfigure}
\usepackage{calc}

\newcommand{\ztwo}{$z\sim2 \hspace{4pt}$}

\newcommand{\ha}{H$\alpha~$}				    
\newcommand{\hb}{H$\beta~$}					    
\newcommand{\nii}{[N{\small II}]}                                       
\newcommand{\sii}{[S{\small II}]}                                       
\newcommand{\oii}{[O{\small II}]}
\newcommand{\oiii}{[O{\small III}]}
\newcommand{\niiha}{\nii/\ha}				    
\newcommand{\siiha}{\sii/\ha}				    
\newcommand{\hahb}{H$\alpha$/H$\beta~$}
\newcommand{\oiiihb}{[O{\small III}]/\hb}
\newcommand{\ro}{r$_{0}~$}
\newcommand{\nel}{n$_{e}~$}

\newcommand{\msun}{M$_{\odot}$} 			   
\newcommand{\mstar}{M$_{*}$} 				    
\newcommand{\mdotout}{$\dot{\rm M}_{\rm out}$}		
\newcommand{\pdotout}{$\dot{\rm p}_{\rm out}$}		
\newcommand{\vout}{v$_{out}$}
\newcommand{\lam}{$\lambda$}
\newcommand{\siggas}{$\Sigma_{gas}$}
\newcommand{\Mw}{M$_{w}$}
\newcommand{\Ew}{E$_{w}$}
\newcommand{\Pw}{P$_{weight}$}
\newcommand{\Pkin}{P$_{kin}$}
\newcommand{\Pturb}{P$_{turb}$}
\newcommand{\Prad}{P$_{rad}$}

\newcommand{\msunyr}{\msun\ yr$^{-1}$}		    
\newcommand{\kms}{km~s$^{-1}$}			              
\newcommand{\msunpc}{\msun pc$^{-2}$}

\newcommand{\lcdm}{$\Lambda${\rm CDM}}		     

\shorttitle{Large-scale outflows from \ztwo galaxy}
\shortauthors{S. F. Newman et al.}

\begin{document}

\title{Shocked Superwinds from the \ztwo  \  Clumpy Star-forming Galaxy, ZC406690\footnotemark[*]}
\author{Sarah F. Newman\footnotemark[1,14], Kristen Shapiro Griffin\footnotemark[2], Reinhard Genzel\footnotemark[1,3,4], Ric Davies\footnotemark[3], Natascha M. F\"orster-Schreiber\footnotemark[3], Linda J. Tacconi\footnotemark[3], Jaron Kurk\footnotemark[3], Stijn Wuyts\footnotemark[3], Shy Genel\footnotemark[3,5], Simon J. Lilly\footnotemark[6], Alvio Renzini\footnotemark[7], Nicolas Bouch\'e\footnotemark[8], Andreas Burkert\footnotemark[9], Giovanni Cresci\footnotemark[10], Peter Buschkamp\footnotemark[3], C. Marcella Carollo\footnotemark[6], Frank Eisenhauer\footnotemark[3], Erin Hicks\footnotemark[11], Dieter Lutz\footnotemark[3], Chiara Mancini\footnotemark[7], Thorsten Naab\footnotemark[12], Yingjie Peng\footnotemark[6] and Daniela Vergani\footnotemark[13]}

\footnotetext[*]{Based on observations obtained at the Very Large Telescope
                (VLT) of the European Southern Observatory, Paranal, Chile
                (ESO Programme IDs
                 183.A-0781, 087.A-0081).}
\footnotetext[1]{Department of Astronomy, Campbell Hall, University of California, Berkeley, CA 94720, USA}
\footnotetext[2]{Aerospace Research Laboratories, Northrop Grumman Aerospace Systems, Redondo Beach, CA 90278, USA}
\footnotetext[3]{Max-Planck-Institut f\"ur extraterrestrische Physik (MPE), Giessenbachstr.1, D-85748 Garching, Germany}
\footnotetext[4]{Department of Physics, Le Conte Hall, University of California, Berkeley, CA 94720, USA}
\footnotetext[5]{Harvard-Smithsonian Center for Astrophysics, 60 Garden Street, Cambridge, MA 02138 USA}
\footnotetext[6]{Institute of Astronomy, Department of Physics, Eidgen\"ossische Technische Hochschule, ETH Z\"urich, CH-8093, Switzerland}
\footnotetext[7]{Osservatorio Astronomico di Padova, Vicolo dellÕOsservatorio 5, Padova, I-35122, Italy}
\footnotetext[8]{Department of Physics and Astronomy, University of California, Santa Barbara, Santa Barbara, CA 93106, USA}
\footnotetext[9]{Universit\"ats-Sternwarte Ludwig-Maximilians-Universit\"at (USM), Scheinerstr. 1, M\"unchen, D-81679, Germany}
\footnotetext[10]{ Istituto Nazionale di AstrofisicaÐOsservatorio Astronomico di Arcetri, Largo Enrico Fermi 5, I Ð 50125 Firenze, Italy}
\footnotetext[11]{Department of Astronomy, University of Washington, Box 351580, U.W., Seattle, WA 98195-1580, USA}
\footnotetext[12]{Max-Planck Institute for Astrophysics, Karl Schwarzschildstrasse 1, D-85748 Garching, Germany}
\footnotetext[13]{INAF Osservatorio Astronomico di Bologna, Via Ranzani 1, 40127 Bologna, Italy}
\footnotetext[14]{email: sfnewman@berkeley.edu}

\begin{abstract}
We have obtained high-resolution data of the \ztwo ring-like, clumpy star-forming galaxy (SFG) ZC406690 using the VLT/SINFONI with AO (in K-band) and in seeing-limited mode (in H- and J-band). Our data includes all of the main strong optical emission lines: \oii, \oiii, H$\alpha$, H$\beta$, \nii, and \sii. We find broad, blueshifted \ha and \oiii $~$ emission line wings in the spectra of the galaxy's massive, star-forming clumps ($\sigma \sim$ 85 \kms) and even broader wings (up to 70$\%$ of the total \ha flux, with $\sigma \sim$ 290 \kms) in regions spatially offset from the clumps by $\sim$ 2 kpc. The broad emission likely originates from large-scale outflows with mass outflow rates from individual clumps that are 1--8x the SFR of the clumps. Based on emission line ratio diagnostics (\niiha and \sii /H$\alpha$) and photoionization and shock models, we find that the emission from the clumps is due to a combination of photoionization from the star-forming regions and shocks generated in the outflowing component, with 5--30$\%$ of the emission deriving from shocks. In terms of the ionization parameter (6x10$^{7}$-Ð10$^{8}$ cm/s, based on both the SFR and the O$_{32}$ ratio), density (local electron densities of 300--1800 cm$^{-3}$ in and around the clumps, and ionized gas column densities of 1200--8000 \msun/pc$^{2}$), and SFR (10--40 \msunyr), these clumps more closely resemble nuclear starburst regions of local ULIRGs and dwarf irregulars than HII regions in local galaxies. However, the star-forming clumps are not located in the nucleus as in local starburst galaxies but instead are situated in a ring several kpc from the center of their high-redshift host galaxy, and have an overall disk-like morphology. The two brightest clumps are quite different in terms of their internal properties, energetics and relative ages, and thus we are given a glimpse at two different stages in the formation and evolution of rapidly star-forming giant clumps at high-z.
\end{abstract}


\section{Introduction} 
Galactic-scale outflows are ubiquitous in local starburst galaxies and in most high-redshift star-forming galaxies (SFGs) \citep{HecLehArm93,LehHec96,Ste+96,Fra+97,Pet+00,Pet+01,Shapley+03,Ade+03,Ade+05,Vei+05,Erb+06,Wei+09,Sat+09,Rub+10,Law+11}. These ÔsuperwindsÕ have a profound impact on the gas content and metallicity of the inter-galactic medium (IGM), injecting enriched, hot and cold gas into the halos and often out of the gravitational potentials of the host galaxies \citep{Mar98,Dev+99,Leh+99,Str+04,RupVeiSan05,Mar05,Ste+10,Stu+11}. Star-formation feedback in the form of these winds is key to understanding galaxy evolution, as they can modulate the star formation activity in gas-rich disks \citep{Kat+96,SomPri99,Cole+00,SprHer03,FinDav08,Ker+09,Genel+10,Dave+11}.

Galactic winds in local starburst galaxies are most easily observed in edge-on systems in optical, near-IR, soft x-ray and radio emission, but outflows have also been detected in face-on galaxies from blue-shifted absorption by dense clouds in the ISM in front of the galactic disks \citep[and references therein]{Vei+05}. While these winds are clearly important for ongoing galaxy evolution and star-formation, little is known about the role of such large-scale outflows in high-redshift galaxies, particularly between z $\sim$ 1--3, when massive local galaxies were undergoing their most powerful epoch of star formation and nuclear activity \citep{Mad+96,Lil+96,Ste+99,Jun+05,Lef+05,Leb+09}.

In the last 15 years, outflows at high redshift have been increasingly observed, most often in Lyman Break Galaxies (LBGs) and QSO absorption line systems \citep{Ste+96,Fra+97,Pet+00,Pet+01,Shapley+03,Ade+03,Ade+05,Vei+05,Wei+09,Rub+10}. Recently, Shapiro et al. (2009) observed broad \ha emission lines from the stacked spectra of 47 \ztwo SFGs, possibly indicative of outflows of several hundred \kms and mass outflow rates comparable to the galaxies' SFRs. In addition, \cite{Ste+10} found interstellar (IS) absorption and Ly$\alpha$ emission from gas spatially offset from $\sim$100 z $\sim$~2--3 LBGs using background galaxies, indicating that most of their sample of high-z galaxies produced metal-enriched galactic outflows. A subset of this sample was further analyzed by \cite{Law+11} with HST/WFC3 imaging who found that outflow characteristics were correlated with rest-frame optical morphology. These previous studies did not have sufficient spatial resolution to resolve the source of the outflows, however. More recently, with high spatial resolution adaptive optics integral-field observations, \cite{Gen+11} have observed powerful outflows emanating directly from massive, M $\sim$ 10$^{9-10}$ \msun, star-forming clumps in five rotationally-supported SFGs at z $\sim$ 2. These clumps are at the high-mass end of the types of clumps observed with HST imaging by \citet{For+11a,For+11b} in \ztwo SFGs and by \citet{ElmElmShe04,Elm+05,ElmElm06,Elm+09a,ElmElm10,Guo+11} in tadpole, chain, and clump-cluster galaxies in the Hubble UDF at z $\sim$~1--5. In this paper, we present deep, follow-up, multi-band observations of one of the galaxies from \cite{Gen+11}, ZC406690.

ZC406690 is part of the ``zC-SINF'' program aimed at studying the spatially-resolved kinematic and star-formation properties of \ztwo galaxies with the SINFONI integral field spectrograph at the ESO Very Large Telescope \citep{Man+11}. The galaxy was originally drawn from the K-band limited sample in the COSMOS field \citep{McCra+10} with the additional ``sBzK'' color criterion, selecting candidate 1.4 \textless~  z \textless~ 2.5 star-forming galaxies \citep{Dad+04}.  Its redshift of 2.19 was confirmed from the zCOSMOS-Deep optical spectroscopic campaign \citep{Lil+07}. The SINFONI seeing-limited and adaptive optics-assisted observations revealed that it is a rotating, clumpy ring with \mstar~$\sim$ 4 x 10$^{10}$ \msun, and M$_{gas}\sim$ 2 x 10$^{11}$ \msun, where M$_{gas}$ is uncertain by a factor of 2--3 \citep{Man+11,Gen+11}. With a UV-derived star-formation rate (SFR$_{UV}$) of 337 \msunyr, ZC406690, falls on the SFR-\mstar main sequence on the high-mass and high-SFR end of the range of SINS and zC-SINF galaxies \citep{For+09,Man+11,Wuy+11a} and the sample from \citet{For+11a,For+11b}. This SFR is in good agreement with the total amount of star formation estimated from \ha corrected for extinction with the best-fit Av from SED modeling and adopting an extra nebular correction of 2.5 (354 \msunyr) \citep[see][]{Cal+00}.
 
\cite{Gen+11} found that ZC406690 is ejecting mass several times faster than it is forming stars (\mdotout $\sim$ 3--8 x SFR), as traced through broad, blue-shifted \ha emission line wings centered on most clumps, indicative of an outflowing warm ionized component. This finding is consistent with previous galaxy-integrated observations of outflows from SFGs at high-z \citep{Ste+10,Pet+00,Wei+09}. With an observed wind outflow velocity of 440 \kms, the brightest and most massive clump in ZC406690 falls very closely on the \vout-\mstar and \vout-SFR relations observed by \cite{Wei+09} for high-z galactic outflows. This suggests that whatever is driving the relation between star-formation and feedback on galactic scales still holds on the kpc-scales of giant clumps. 

This compelling evidence for massive, galactic scale outflows originating in $\sim$1 kpc clumps has motivated us to follow-up on these original observations. We exploit here the deep SINFONI+AO data to create individual spectra of not just the clumps, but of the surrounding regions, whose signal is dominated by the outflowing component projected onto the disk. With excellent S/N we are able to create emission line maps of \nii\lam6584 and \sii\lam6716 as well as the broad and narrow components of the \ha emission line, enabling us to probe the spatially resolved structure and density of the wind. With new seeing-limited J and H band data of the galaxy, we obtain emission line fluxes of [OIII]\lam5007,4959, H$\beta$, and [OII]\lam3726,3729. This new data enables us to put the clumps in the BPT diagram \citep{BPT81,VeiOst87}, along with local SFGs and starbursts as well as other high-z SFGs. We can then compare the ISM properties (ionization state, SFR, density, metallicity, star-formation history) of spatially-resolved high-z star-forming clumps with those of different types of local star-forming regions to better understand the conditions required for these massive outflows and how the clumps fit into the overall picture of galaxy evolution. 

In section 2, we outline the observations used as well as our reduction and analysis techniques. In section 3, we present the new spectra of individual regions, emission-line maps, and shock-diagnostic diagrams based on more detailed examination of the K-band data and the new H- and J-band data. In section 4, we compare the results to those of local and high-z star forming galaxies and discuss the implications. In section 5, we summarize our findings.  Throughout this paper, we adopt a \lcdm~ cosmology with $\Omega_{m}$ = 0.27, $\Omega_{b}$ = 0.046, and H$_{0}$ = 70kms$^{-1}$ Mpc$^{-1}$ \citep[WMAP7,][]{Kom+11}, as well as a \cite{Cha03} initial stellar mass function (IMF).

\section{Observations, Data Reduction and Analysis Techniques}
\subsection{Observations and Data Reduction}

We observed ZC406690 with SINFONI/VLT integral field unit spectroscopy \citep{Eis+03,Bon+04} using natural guide star adaptive optics (AO) in K band for 10 hours to achieve a spatial resolution of 0.22'' (1.85 kpc) FWHM (as presented in \cite{Gen+11}). Each observation block (OB) consisted of an ``O-S-O-O-S-O'' sequence, with 4 on-source exposures and 2 others taken on empty `sky' positions about 20 arcsec away from the target.  For each on-source frame, the object's position was varied around the center by $\pm$ 0.4'' to avoid redundancy. Off-source ``sky'' frames were necessary because the source fills a significant fraction of the 3.2'' x 3.2'' FOV at that pixel scale. Individual exposure times were of 600s, giving 2400s on-source per OB, and thus a total on-source integration time of 10h (for 15 OBs). The resulting cube has 0.05'' spatial pixels and 0.000245 $\mu$m spectral pixels.  

We also observed the galaxy in seeing-limited mode in J-- (100 minutes) and H--band (120 minutes), obtaining $\sim$0.6'' FWHM with a pixel size of 0.125'' x 0.000195 $\mu$m. With this pixel scale, the larger FOV of 8'' x 8'' allowed an on-source dithering strategy, such that each OB consisted of six on-source exposures, with the source positioned alternatively in one or the
other half of the FOV, with additional small dithering of $\sim$1/10 of the FOV to avoid redundant positions. Individual exposure times were all of 600s, giving 3600s on-source per OB, and a total on-source integration time of 2h per band (for 2 OBs each). Note that we only use 100 minutes of integration time in the reduced J--band cube, as the first two exposures had faulty pointings.

For all AO and no-AO OBs, images of the acquisition star taken before the science exposures were used to monitor the positioning and the PSF for the OBs. Standard stars of O or early-B types were observed each night, at similar airmass as the science data, and used to correct for the
telluric transmission and to provide the absolute flux calibration.

For the data reduction, we used the software package SPRED \citep{Sch+04} and custom routines for optimizing the background/OH airglow subtraction. For combining the individual exposures, we determined the relative offsets between frames using the known offsets for frames within individual OBs and using the position of the aquisition star, for frames in different OBs. The effective PSF FWHM was determined from the combined PSF data sets associated with all OBs in a given band (fitting a 2D gaussian profile to the image made by averaging over all spectral channels). For more on the data reduction procedure, see \cite{For+09,Man+11}.

We calculate the SFR from the \ha luminosity using the \cite{Ken98} conversion adjusted to our adopted \cite{Cha03} IMF, and corrected for extinction.  The extinction correction is described in Section 3.1.

\subsection{Emission Line Maps}

With the reduced K-band data cube, we create H$\alpha$, \nii\lam6584, and \sii\lam6716,6731 maps by simultaneously fitting Gaussian components to each of the emission lines, holding the nebular \nii\lam6584/6548 ratio fixed at the atomic value (3.28) and using the kinematics (line center, dispersion) derived from \ha and errors determined from the noise cube. Prior to fitting the emission lines, we smooth the data by 3 spatial pixels in each direction and 3 spectral pixels (5x3 for the \sii$~$ map) to boost the S/N for the fainter nebular lines. Note that we only include a map for \sii\lam6716, since \sii\lam6731 is strongly affected by an OH sky line in some spatial elements. For all line maps, we show only pixels with S/N greater than 3$\sigma$. 

We use the same reduction procedure for the H- and J-band cubes, but use a different technique to produce the emission-line maps. Since the S/N is not high enough to fit each line individually for each spatial pixel, the maps were obtained by spectrally integrating over the full width of each line. We further increase the S/N of the maps by smoothing over 3x3 spatial pixels. After the smoothing, all three maps were clipped by a S/N ratio of 3 and were overlayed with \oiii$~$ intensity contours.

Broad and narrow \ha emission line maps are created by fitting a two-component Gaussian to the emission lines. The resulting broad \ha map is used to visualize the trend in broad emission surrounding one of the clumps, but the map itself has very low S/N, so we generate an alternate map for display purposes. This other broad \ha map is created by spectrally integrating over the wings of the \ha line for the velocity intervals [-430, -140 \kms] and [280, 570 \kms] around the centroid of the narrow component. The two maps of the broad \ha emission line are qualitatively similar. 

\subsection{Extraction of Spectra from Individual Regions}

We create spectra of individual regions from our K-band data (Figures 1 and 2) by averaging the spectra over both narrow-line (clump) and broad-line (wind) regions.  We select the ÔwindÕ component based on the region of elevated \niiha around the clumps (see section 3.3). We fit the broad emission lines using two-component Gaussian fits with the velocity and width of all lines held equal to that of the \ha line. In some cases, the lines were fit equally well with a single Gaussian, which we assumed to be the case if the difference in $\chi^2$ values over the \ha and \nii$~$ region of the spectrum between the one- and two-component fits was less than 10\%. For those spectra (clumps C and D), we use the single Gaussian fit. We find that the emission line profiles in all regions are well described by either one or two Gaussians. We report the relative contribution to the flux by the narrow and broad components in the next section and Table 1. We derive the error for the fit parameters (velocity dispersion, velocity, etc) by generating 1000 monte carlo realizations of the fit and taking the standard deviation for each parameter.

For the H- and J-band spatially integrated spectra, we fit each of the \oiii\lam4959,5007, H$\beta$, and \oii\lam3726,3729 lines with a single Gaussian component because these lines have too low S/N to fit the underlying broad component. For H-band, we fit all of the lines using the kinematics of the \oiii\lam5007 line, and for J-band we constrain the two \oii$~$ lines to have the same kinematics as \ha and fix their wavelength separation to the known value, but allow their relative intensities to vary. For emission line ratios that contain lines from multiple bands, we employ an extinction correction as outlined in Section 3.1 with an E(B-V) as determined from \hahb for clump A. In addition, for all ratios (except \niiha and \sii/H$\alpha$), we use emission line fluxes derived from single-component Gaussian fits. When comparing K-band (higher spatial resolution) and H-band emission lines, we resample the K-band cube to match the H-band cube, and convolve the new K-band cube with the H-band PSF. The H- and J-band cubes have the same spatial and spectral resolution. We therefore use the exact same spatial pixels when extracting spectra of individual regions from H, K and J bands.

To create the \sii$~$ line profiles (Figure 2) and ratios, we generate a Òvelocity-shiftedÓ cube to increase the S/N. This is done by shifting the spectrum of each pixel based on the measured \ha velocity such that all pixels have their \ha peak at the same wavelength. This method has been shown to increase the S/N of region-integrated spectra \citep{Sha+09}. To negate the effects of an OH sky line that lies near the \sii\lam6731 emission line for parts of the galaxy, we create clump A and clump B spectra that have maximal S/N ratios in the wavelength range of this line. To this end, we define the spatial regions of the clumps by only adding spatial pixels to the spectra such that each additional pixel increases the S/N over a given threshold. Essentially, we only add pixels with a \sii\lam6731 S/N ratio  $\ge$ 4. In this way, we only use measurements of the \sii\lam6731 line in pixels that are minimally affected by the OH sky line. In order to maximize the S/N, these regions must be quite extended, so to get \sii\lam6731 measurements for smaller areas within these larger regions (i.e. for the broad regions surrounding the clumps), we assume the measured \sii\lam6716/\sii\lam6731 ratio within the larger regions is constant. We can then estimate \sii\lam6731 fluxes based on the measured \sii\lam6716 flux in the smaller regions and the measured \sii$~$ ratio from the corresponding larger region.

The \hb line is obstructed by an OH atmospheric line for about half of the galaxy. Therefore, to obtain an \hb flux estimate for clump B (which falls in this right half), we assume clump B follows the extinction of the dense, clump gas and has the same observed \hahb ratio as clump A. We obtain an \hb flux using this ratio and the \ha measurement for clump B, although this estimate is much more uncertain than the clump A \hb flux.

\section{Results} 
\subsection{The Clumps are More Highly Extincted than the Surrounding Gas}

The extinction correction used to calculate the SFR is based on the best-fit visual extinction derived from stellar population modeling of the UV to mid-IR spectral energy distribution (SED) presented by \cite{Man+11}, for a \cite{Cal+00} reddening law. Following \cite{Cal+00}, we assume that the \ha line emission is more attenuated than the bulk of stellar continuum light and thus correct the observed \ha luminosity for A$_{V(nebular)}$ = A$_{V(SED)}$ x 2.5 \citep[see also][]{For+09,Man+11,Wuy+11a}. 

We also calculate the extinction for clump A and the left-half of the galaxy (which is unaffected by an OH sky line near the \hb line) based on the \hahb ratio and using the \cite{Cal+00} reddening law, assuming an intrinsic ratio of 2.86 \citep{Ost89} appropriate for an electron density of 100 cm$^{-3}$ and a temperature of 10$^{4}$ K. We find that the SED-based A$_{V}$ (described above) assumed in \cite{Gen+11} and \cite{Man+11} of 1.3 $\pm$ 0.5 (E(B-V) = 0.33 $\pm$ 0.1) is consistent with the diffuse emission from the galaxy (as calculated with \hahb from the left-half, A$_{V}$ = 1.5 $\pm$ 0.8, E(B-V) = 0.37 $\pm$ 0.18) and that the nebular emission in clump A is more highly extincted (A$_{V}$ = 2.0 $\pm$ 0.8, E(B-V) = 0.49 $\pm$ 0.17), as one would expect for a star-forming region. All of the forthcoming calculations use the clump \ha flux as calculated and corrected for extinction in \cite{Gen+11}. If we were to apply an extinction correction based on the \hahb measurement for clump A, the main clump properties (L$_{H\alpha}$, SFR, \mdotout, \siggas) would be increased by about 50\%, and therefore the calculations made in Section 4 based on these quantities would be scaled accordingly. This relatively small factor will not affect any of the conclusions made in this paper.

\subsection{Spectra of the Clumps and their Surroundings are Indicative of Localized Superwinds}

We explore the properties of the star-forming clumps by creating individual spectra of both the clumps and their surrounding regions. In the regions surrounding the clumps, we find broader \ha emission lines and enhanced \niiha and \siiha ratios as compared to the clump regions. Enhancement of these emission line ratios could be due to several physical processes including the presence of shocks in a large-scale outflow or an AGN. 

\begin{figure}[tb]
\centerline{
\includegraphics[width=3in]{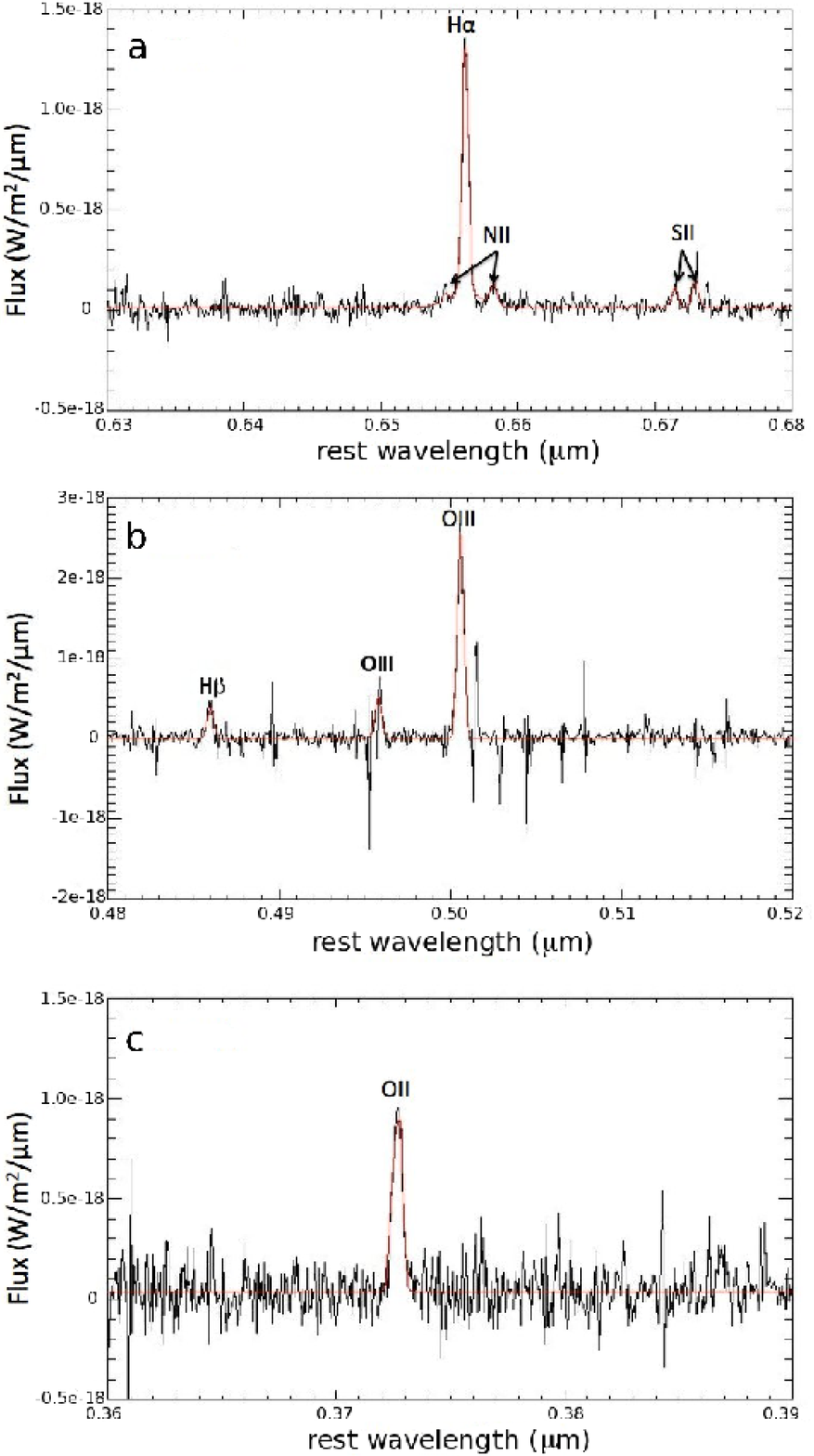}}
\caption{Panels a, b and c show K, H and J band spectra of clump A with the relevant emission lines labeled, respectively. The black line represents the data, while the red line shows our best fit. The wavelength axis is in rest-frame units.}
\end{figure}

\begin{figure*}
\centerline{
\includegraphics[width=7in]{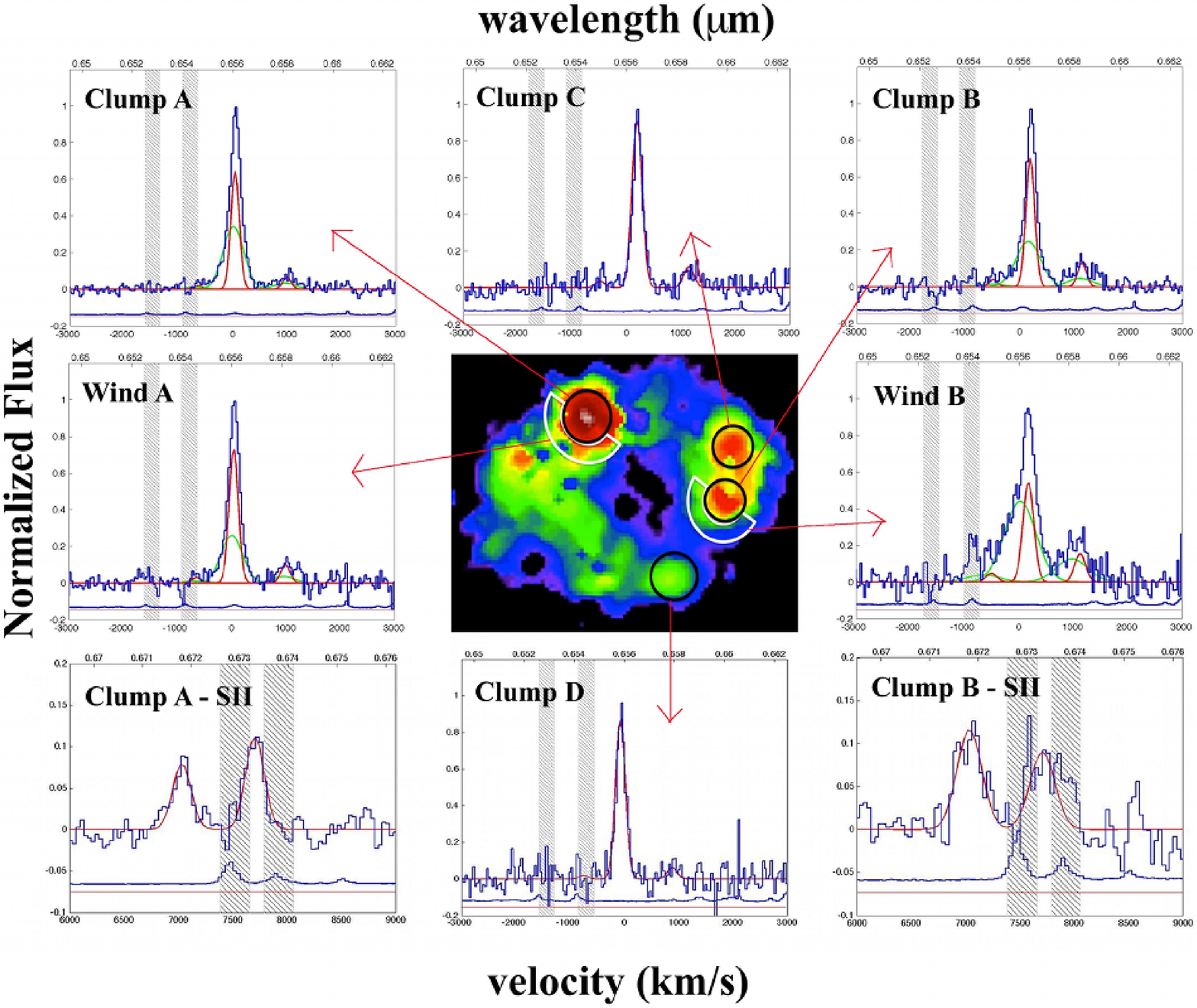}}
\caption{Emission line profiles and best fit to \ha and \nii\lam6584 features for six regions of ZC406690: clump A, clump A `wind',  clump B, clump B `wind' and clumps C and D.  Included with these spectra are one- or two-component Gaussian fits. For the two-component fits, the narrow component is shown in red and the broad in green. For the spectra with one-component fits, these are shown in red. The errors (1$\sigma$ rms) are shown below in blue. Clumps A and B and their winds are well fit by two Gaussians, while clumps C and D can be well fit by one component. The lower right and left panels of Figure 2 compare the spectra and one-component Gaussian fits of the \sii$~$ emission lines for clumps A and B, with regions selected by the method described in 2.3. For clump A, \sii\lam6716/\sii\lam6731 = 0.7 +/- 0.1, while for clump B, = 1.2 +/- 0.3, corresponding to electron densities of 1800 $\pm$ 1000 and 290 $\pm$ 300 cm$^{-3}$, respectively \citep{Ost89}. The grey hatched lines in all figures show the wavelength range of strong sky OH emission features. The central panel is an \ha map of ZC406690, with each region outlined.}
\end{figure*}

In this case, however, it is very unlikely that these large line ratios are due to an AGN. First, the region in which we find these large \nii/H$\alpha$ and \siiha ratios is extended, unlike an AGN point source. Second, these regions are located around clumps in the disk (or ring) of the galaxy, far from the center, where one would expect to find an AGN. The velocity field of the ring is also very regular and characteristic of face-on rotation, unlike what would be expected if clump A and clump B are the nuclei of merging galaxies. Third, the value of the elevated \niiha ratio in this region (0.3-0.5) is much less than those expected for AGN (at least 0.6-0.8). Fourth, we find that the \nii$~$ emission is well fit by both a broad and a narrow component (see later in this section), which is inconsistent with classical broad-line regions around AGN. Finally, there is no evidence of an AGN from the ZCOSMOS optical spectra of this galaxy \citep{Man+11}. 

We further discuss the possibility of shocks in Section 3.3, but even without the emission line ratios, the broad emission line wings (up to 800 \kms) in the spectra of the regions surrounding the clumps seem to indicate the presence of outflowing material. We therefore call these regions the clump A and clump B `wind' regions.

In Figure 1, we show K--, H--, and J--band spectra from Clump A (the brightest and most massive clump) along with our best fit model. In Figure 2, we show emission line profiles of \ha and \nii\lam6584 for the four clumps in ZC406690 and the wind regions surrounding clumps A and B. Clumps A and B and the surrounding regions are well fit by two Gaussian profiles, with narrow components with FWHM = 200$\pm$8 \kms and FWHM = 210$\pm$9 \kms and underlying broad components with FWHM = 470$\pm$28 \kms and FWHM = 630$\pm$82 \kms, which we ascribe to outflowing material. On the other hand, clumps C and D can be well fit by one component with FWHM = 92$\pm$2 \kms and 89$\pm$2 \kms, suggesting little or no outflow from these clumps. The clump B wind spectrum is the most extreme example of a localized superwind with a very broad blue line wing, with the edge of the wing extending out to roughly 800 \kms from line center. In addition, the centroid of the broad component is blueshifted by 150$\pm$16 \kms from the narrow. This is consistent with the signature of a powerful outflow, as one would expect to see the majority of the outflowing material as blueshifted if the clumps are heavily obscured. The redshifted emission would fall behind the plane of the galaxy, and much of this emission would be extincted as it passes through the clumps as suggested by our \hahb measurements.

The fraction of \ha emission that comes from a broad-velocity component varies substantially for different regions in the galaxy. For the galaxy as a whole, broad emission accounts for roughly 33\% $\pm$ 2\% of the total \ha flux. For clumps A and B and wind A, the broad component accounts for about 50\% $\pm$ 6\% of the total \ha flux, while for wind B, broad emission comprises 70\% $\pm$ 3\% of the \ha flux. In all of the clump and wind regions, the narrow component likely derives from star-formation in the disk, while the broad component may come from a combination of turbulence and outflowing gas. In the case of wind B, the broad velocity-component peak wavelength is blueshifted slightly from the narrow-velocity component peak, which indicates that a large fraction of the \ha flux from this region comes from an outflowing wind. The emission line profile for \oiii$~$ is very similar to that for H$\alpha$, such that the clump A \oiii$~$ line is slightly less broad than H$\alpha$, but for clump B, both lines are equally broad, with the asymmetric blue line wing also present for \oiii. However, we note that the aforementioned comparisons are done by eye since the S/N for the H-band spectrum for each clump is too low to perform a rigorous two-component fit. Since \oiii$~$ and \oii$~$ likely have similar kinematics, our choice of using \ha kinematics to fit the \oii$~$ lines is thus supported. 

We derive the local electron density of the clumps using the ratios of the \sii$~$ and \oii$~$ doublets. The lower right and left panels of Figure 2 compare the spectra and one-component Gaussian fits of the \sii$~$ emission lines for clumps A and B, with regions selected by the method described in 2.3. For clump A, \sii\lam6716/\sii\lam6731 = 0.7 +/- 0.1, while for clump B, = 1.2 +/- 0.3, corresponding to electron densities of 1800$\pm$1000 and 290$\pm$300 cm$^{-3}$, respectively \citep{Ost89}. These densities likely come from a combination of both the star-forming regions in the disk and high-density shocked filaments or clouds in the outflow, from which the emission lines originate. Since the clump+wind B region has a larger broad H$\alpha$ flux fraction than the clump+wind A region, we would expect that more of the \sii$~$ emission for clump B derives from the outflow than for clump A. Therefore, the observed clump A electron density of 1800 cm$^{-3}$ is probably a better estimate for the dense clump environment, while the clump B density of 290 cm$^{-3}$ is likely an average of clump gas and less dense outflowing cloud gas. The clump B estimate is similar to the value assumed by \cite{Kew+01b} in their starburst models (350 cm$^{-3}$), which is based on the average density of individual HII regions from their sample observed with a 1 kpc slit.

Densities calculated from the OII doublet are much more uncertain (as the lines are mostly blended together in our spectra), and are only consistent (given the uncertainty) with the \sii$~$ derived densities for clump B (see Table 2). This discrepancy could could be due to the two tracers probing different regions in the wind. The 1$\sigma$ ranges quoted in Table 2 for both the \sii-- and \oii--derived densities are obtained by adding and subtracting the uncertainties derived from the emission line fits to the ratio value and finding the corresponding electron densities from \cite{Ost89}.

We interpret the wind B region as the projection of an outflowing component onto a low-inclination disk ($\sim20^{o}$ ($^{+10}_{-5}$), as estimated in \cite{Gen+11}). Since regions adjacent to the clumps would have much lower \ha luminosity from star formation than the clumps, the broad emission from the winds can account for much more of the total signal. Furthermore, if we assume the outflow has a biconic geometry \citep[and references therein]{Vei+05}, then the distant parts of the outflow will have a larger spatial extent than the clump, further separating this ``wind'' region from the clump in projection. This notion of viewing the wind in projection is consistent with the presence of less of an offset broad component for clump A, as it is likely younger than clump B (see section 4.2) and therefore has had less time to produce an extended wind. The relative outflow velocities from the two clumps further supports this view (with v$_{out,B} >> $ v$_{out,A}$), as most outflow models \citep{HecArmMil90,Vei+05} predict increasing wind velocity with distance from the galactic plane.

We can use this model of a projected wind to learn about the geometry and structure of the outflow. The offset between clump B and wind B is $\sim$ 1.8 kpc, so with an inclination of 20$^{o}$, the average distance of the line-emitting outflow from the disk is $\sim$ 5.3 kpc. Given that the broad component of wind B is blueshifted by $\sim$ 175$\pm$20 \kms from the narrow component of clump B, then the average timescale of the line-emitting part of the wind is 30 Myr. It is likely that the gas in clump B has been forming stars for much longer than this given its average metallicity (see Table 2), but we are only observing the portion of the outflow coming from dense, hot filaments in the wind, and the older parts of the outflow may be too diffuse or faint to observe.

\subsection{Broad Emission is Spatially Correlated with Shock-like Emission Line Ratios}

\begin{figure*}
\centerline{
\includegraphics[width=7in]{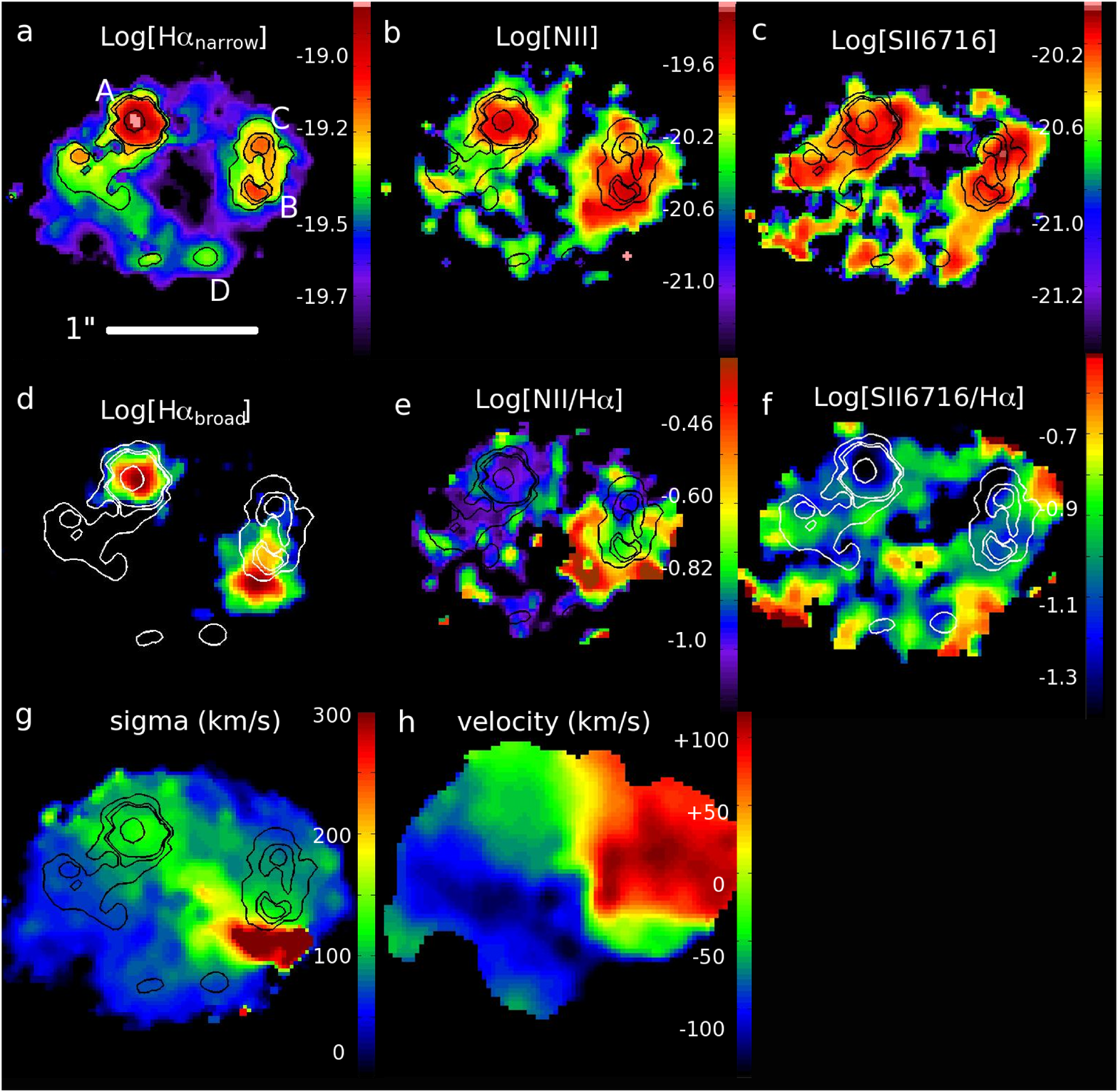}}
\caption{Spatially-resolved emission line maps for (a) log[H$\alpha$ narrow], (b) log[\nii\lam6584], (c) log[\sii\lam6716], (d) log[\ha broad], (e) log[\nii/H$\alpha$], (f) log[\sii\lam6716/H$\alpha$], (g) the \ha dispersion and (h) the \ha velocity field. The top three panels show the velocity-integrated flux and have a S/N per pixel (for the un-resampled maps with 0.05'' pixels) near the clumps of 100-200, 10-40, and 10-40, respectively. \sii\lam6731 is not shown since it is affected by an OH sky line. All of the maps are rebinned by two pixels and smoothed by 3 pixels for the purpose of presentation and are overlaid with \ha intensity contours to highlight the locations of clumps A, B, C and D (labeled in panel a). The rebinning does not change the overall appearance of the maps as the S/N is very high, but merely accentuates features that are already present before rebinning. The spatial resolution of these maps is 0.2''. From panels (d), (e), and (f), we find that broad emission surrounding clump B is spatially correlated with shock-like \niiha and \siiha ratios, indicating an energetic wind coming from the clump.}
\end{figure*}

Figure 3 shows spatially-resolved emission line maps for H$\alpha$, \nii\lam6584 and \sii\lam6716, which illustrate the effect of the outflowing gas on the region surrounding the clumps. The peak in \nii\lam6584 emission is located near the peak of \ha emission over the clumps but the \nii$~$ emission is more extended around clump B than the \ha emission (by $\sim$1 kpc). This effect is seen even better in the ratio map of \niiha located in panel e of Figure 3. There is a semi-circle like structure centered on clump B that peaks in \niiha over the broad-line region, with an enhancement of about 80\% from the clump value (see Table 1). The same effect is seen around clump A with a smaller enhancement. Errors for the different regions are over an order of magnitude lower, such that the uncertainty $\sim$ 2-5\% on and near the clumps. These enhanced values of the \niiha ratio can be indicative of both higher metallicity and evidence of shocks and AGN \citep[e.g.][]{DopSut95}. We previously discussed why an AGN explanation is unlikely in Section 3.2.

We can somewhat disentangle the effect of metallicity and shocks on the \niiha ratio by examining other emission line ratios and comparing the data to shock and photoionization models from the literature. Both the \sii\lam6716+6731/\ha and \oiiihb ratios are much more sensitive to the ionization parameter (and thus shocks) than to metallicity, in the metallicity regime of the clumps.

The \sii\lam6716 map (panel c) shows that much of the emission is located on or near the clumps, although it does not follow the \ha emission very closely. Panel f shows a map of \sii\lam6716/H$\alpha$, which serves as a lower limit for \sii\lam6716+6731/\ha which is omitted due to the aforementioned OH sky lines. Even with this underestimate, we see slightly higher ratios surrounding the clumps than over their centers (0.059 vs. 0.087 for clump A and 0.11 vs. 0.14 for clump B). This is consistent with the notion that the larger \niiha ratios are caused by shocks (and possibly also photoionization from star formation) as opposed to sharp metallicity gradients. The typical uncertainty for the \sii\lam6716/\ha ratio near the clumps is 5-10\%, although the S/N drops off sharply away from the clumps. 

It is possible that there is some variation in the \sii\lam6716/6731 ratio between the clump and wind regions that we are unable to measure because of the OH sky lines, and we have considered the scenario in which this variation could compensate for the gradient of \sii\lam6716/\ha from the clump to wind regions, thereby erasing the overall \siiha ratio gradient. However, if we calculate the variation in the \sii$~$ ratio required to erase the \siiha gradient and assume the clump region has the observed value of the ratio, we get implausible values for the \sii$~$ ratio of the winds, 1.6 for the clump A wind and 2.5 for the clump B wind, which are inconsistent with the collisional de-excitation model at 10$^4$ K from \cite{Ost89} . Thus the gradient in \sii\lam6716/\ha between the clump and wind regions implies a similar gradient in \sii\lam6716+6731/H$\alpha$, which cannot be washed out by density variations.

This region of enhanced emission line ratios surrounding clump B is spatially correlated with the bulk of the broad \ha emission, which we ascribe to an outflowing component. From the broad \ha map (Figure 3 panel d), it is evident that most of the broad emission comes from directly over clump A and to the SE of clump B, coinciding with the enhanced \niiha and \siiha ratios. This broad region near clump B, from which we extracted the `wind B' spectrum, not surprisingly also shows up as a peak in the velocity dispersion in panel g, and is blueshifted relative to clump B, as seen in the velocity map (panel h). The latter statement is confirmed by examining the line centers of the narrow components for clump B and wind B. The kinematic maps are derived by fitting a single Gaussian component to the \ha line (for more on this see \cite{Gen+11}). Thus, if active star formation in the clumps produces a large-scale, energetic outflow, this could explain both the broad emission lines and the enhanced emission line ratios in this region.

In order to visualize the emission line gradients around clump B and understand their errors, we show the average flux across a swath of the emission line maps with a slit width of 0.2'' in Figure 4. This figure shows that there is a relative increase in emission for both \nii, \sii$~$ and \ha broad relative to \ha narrow around clump B with errors from the noise cube that are much smaller than these gradients, and thus the emission line gradients seen in Figure 3 near clump B are real. While the errors increase dramatically for \siiha near the edge of Figure 4, the gradient is seen clearly near clump B where the S/N is relatively high. The \ha broad emission is calculated as described in Section 2.2 using a two-component fit to the \ha line.

\begin{figure}
\centerline{
\includegraphics[width=3in]{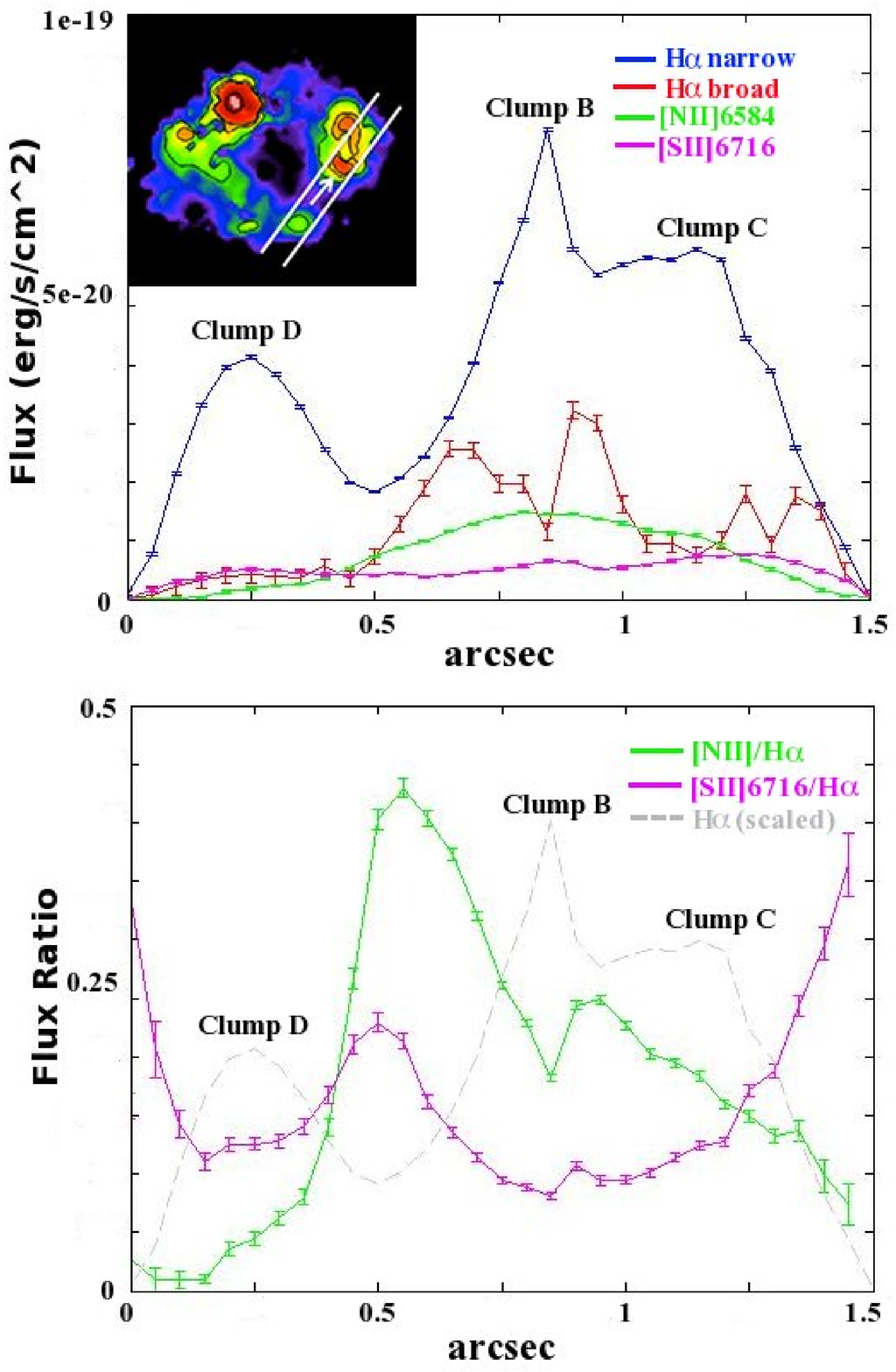}}
\caption{Averaged flux of several emission lines and ratios along a four pixel-wide (0.2'') swath of the galaxy moving from clump D to B to C. (a) Average flux of \ha narrow, \ha broad, \nii\lam6584 and \sii\lam6716, along with error bars from the noise cube. (b) \niiha and \sii\lam6716/\ha and error bars from the noise cube with \ha narrow scaled in the background for reference. There is a relative increase in emission for both \nii, \sii$~$ and \ha broad relative to \ha narrow around clump B and the errors are much smaller than these gradients. For \nii/H$\alpha$, \siiha and \ha broad, the gradients from the peak of emission (or the peak of the ratio) to the minimum (near the clump B center) are over 10$\sigma$. The inset in the upper left corner of (a) shows the width, location and direction of the swath.}
\end{figure}

In Figure 5, we show the emission line maps for \oiii, H$\beta$, and \oii. As is clear from the first and last panels, the \oiii$~$ and \oii$~$ emission is coincident with the location of the clumps. Even with the final smoothing (see Section 2.2), the S/N in individual pixels for \hb$~$ is very low, and thus only a small region near clump A remains after the S/N = 3 clipping. Fortunately, the \hb$~$ line in the integrated spectrum for clump A has a S/N ratio $\sim$ 5, and we are able to use this to derive \oiiihb ratios for the clumps.

\begin{figure}
\centerline{
\includegraphics[width=2.7in]{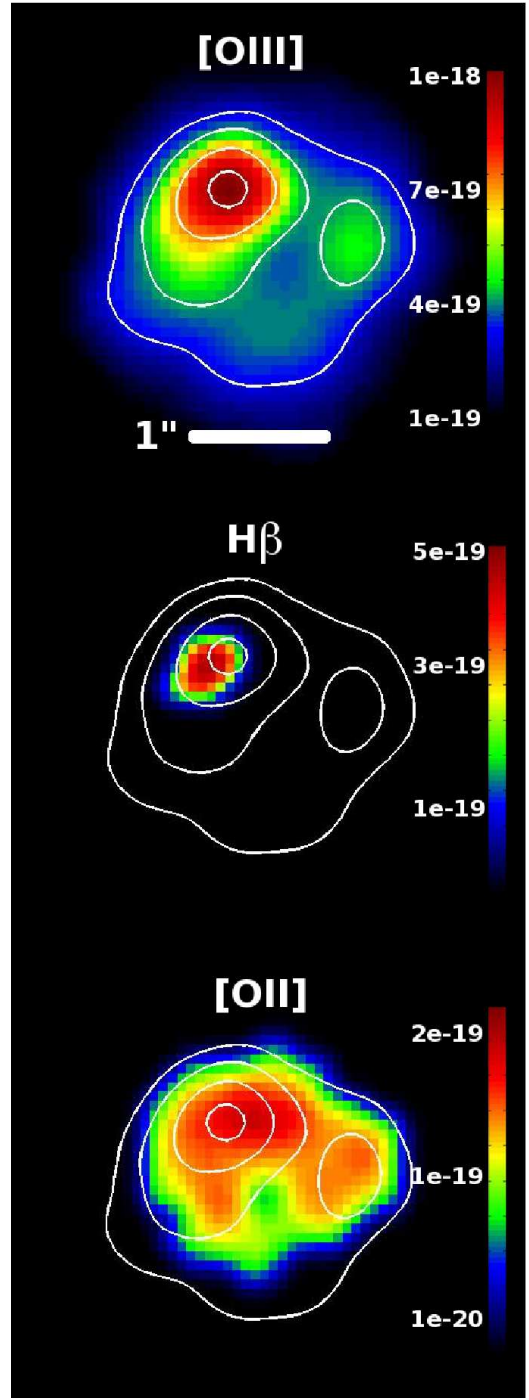}}
\caption{Spatially-resolved emission line maps for (top to bottom) \oiii, H$\beta$, and \oii. The maps have been smoothed by 3x3 pixels to enhance the S/N and then clipped by S/N = 3. The maps were then rebinned by 2 spatial pixels and smoothed again by 3 pixels for the purpose of presentation. The white contours trace the \oiii$~$ flux. The un-resampled cubes have 0.125'' pixels and a PSF FWHM of 0.6''. }
\end{figure}

Figure 6 shows the BPT diagram \citep{BPT81} with SDSS data from the latest DR7 release \citep{Ade+09} along with the data for clumps A and B, to compare the latter to local star-forming galaxies and other high-z galaxies. Since the clump A and B data points use both high-resolution K- and low-resolution H-band data, we extract larger regions, which contain both the `clump' and `wind' regions. We are unable to get a direct measurement of the \hb line for clump B, as it is shifted onto an OH sky line in that part of the galaxy, but we estimate the clump B \hb value assuming a constant \hahb ratio for the clump regions (\hahb = 5.1 as measured from clump A). The position of the clump B point is therefore much more uncertain, and the lower error bar for \oiiihb extends off the figure. The errors come from the uncertainty of the fits to the emission lines. Clumps A and B fall in the ÒcompositeÓ region of the BPT diagram that is generally occupied by starburst and high-z galaxies, which are generally found to have higher SFRs, densities and ionization parameters than normal SFGs and HII regions \citep{Shapley+05,Erb+06,Liu+08,Bri+08}. Their offset could also be due to shocks, AGN, or a different N/O ratio. As the clumps occupy the same parameter space as these galaxies, it appears that their star-forming regions more closely resemble those of starbursting galaxies than of HII regions in normal local SFGs. 

\begin{figure}
\centerline{
\includegraphics[width=3.5in]{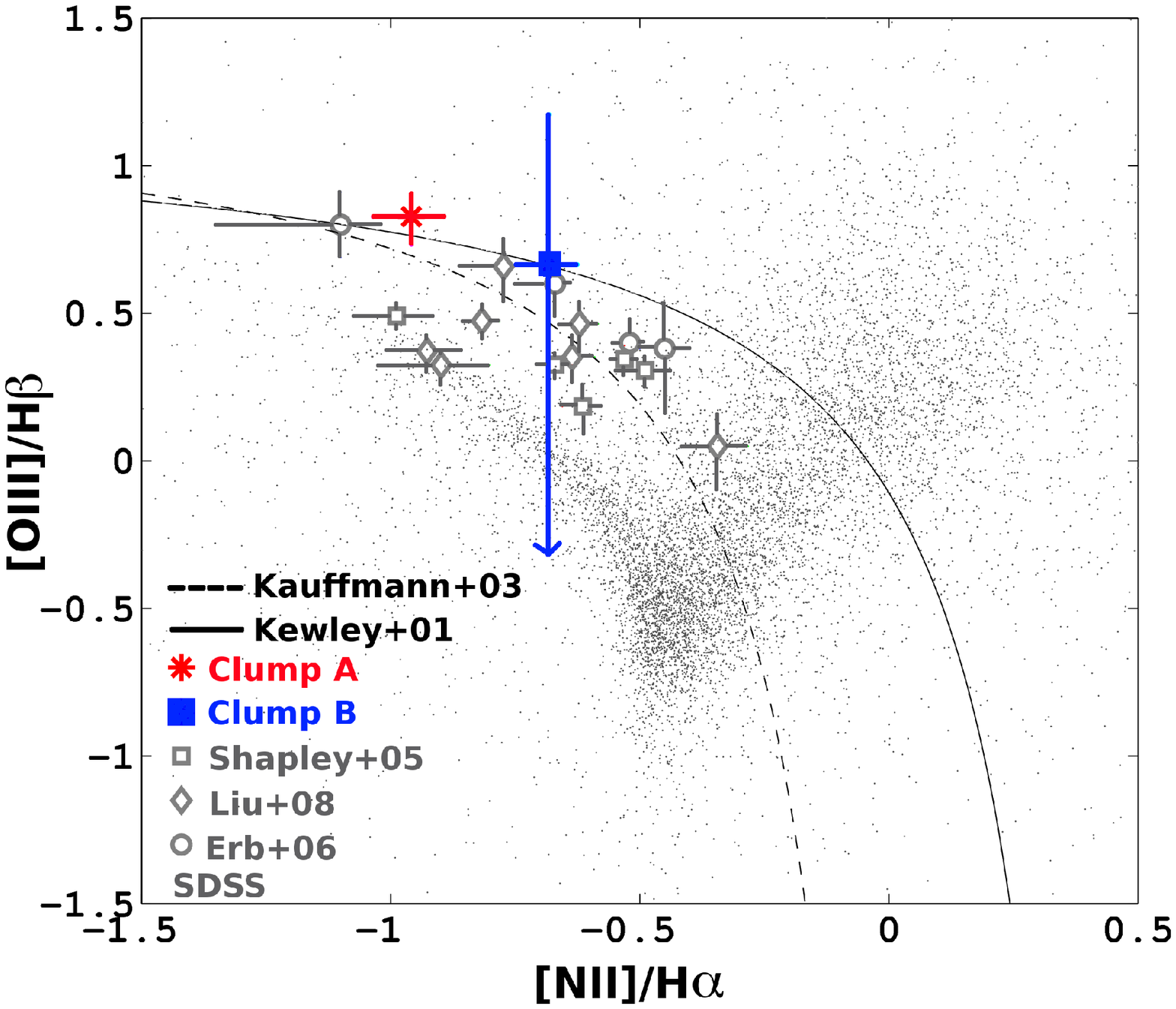}}
\caption{The BPT diagram including data from several high-z galaxies along with the clumps from ZC406690. Clumps A and B are shown as a red asterisk and blue square, respectively. The lower error bar for clump B extends beyond the figure boundaries. Other high-z data is shown as grey squares \citep{Shapley+05}, circles \citep{Erb+06}, and diamonds \citep{Liu+08}. A subset of SDSS galaxies are shown in grey with the empirical AGN/HII boundary as the dashed line from \cite{Kau+03} and the theoretical boundary as the solid line from \cite{Kew+01a}. The clump data points fall in the region offset from the main HII branch that is also occupied by other high-z galaxies and many local starbursts \citep[as shown by][]{Liu+08,Bri+08}.}
\end{figure}

In order to differentiate the effects of metallicity, ionization parameter, and the contribution of shocks, we compare our data to photoionization and shock models from \cite{Ric+11} and photoionization models from \cite{Lev+10}. These photoionization models are similar in many respects: they are both created with the codes Starburst99 \citep{Lei+99} and Mappings III \citep{SutDop93}, use the same evolutionary tracks and both assume continuous star-formation histories, but they also differ somewhat in terms of model atmosphere, geometry and IMF. Although the \cite{Ric+11} shock models are for slow shocks (up to 200 \kms), they sample a large portion of the parameter space (in terms of metallicity and ionization parameter) which fits our observations, and the exact velocity of the shocks do not qualitatively effect our conclusions since there is a much bigger difference between photoionization and shock models than slow vs. fast shocks in terms of emission line ratios. While evidence for shocks indicates the presence of an outflow, emission from both shocks and photoionization could be present in a wind.

Figure 7 shows our data from the clump and wind regions as well as the photoionization and shock models in the \niiha vs. \siiha plane (with \siiha calculated as described in section 2.3), with varying metallicity, ionization parameter and shock velocity. We are able to get data for both the clump and wind regions here as we are only using the high-resolution K-band data while we could only get overall clump regions with lower resolution for Figure 6 since we used lower-resolution H-band data in addition to K-band. Thus the clump regions in Figures 6 and 7 cover different spatial areas and therefore they have emission line ratios with slightly different values. We also note that, as for Figure 6, the error bars come from the uncertainty in the emission line fits and are therefore larger than the error bars shown in Figure 4, which are derived from the noise cube. 

The two photoionization models shown in Figure 7 \citep{Ric+11,Lev+10} are roughly the same for [O/H] = 8.39, however, the [O/H] = 8.69 line is offset to much larger values of the \niiha ratio for the \cite{Lev+10} model. This is likely due to one of the differences between the models mentioned above. At first glance, the data points seem to be well fit by the photoionization-only models. Both wind regions appear to fall near the same metallicity line as their respective clumps, but offset to higher emission line ratios, indicating a lower ionization parameter. However, we can independently calculate the ionization parameter for the clumps, and compare that to the ionization parameter implied by the photoionization models.

\begin{figure}
\centerline{
\includegraphics[width=3.5in]{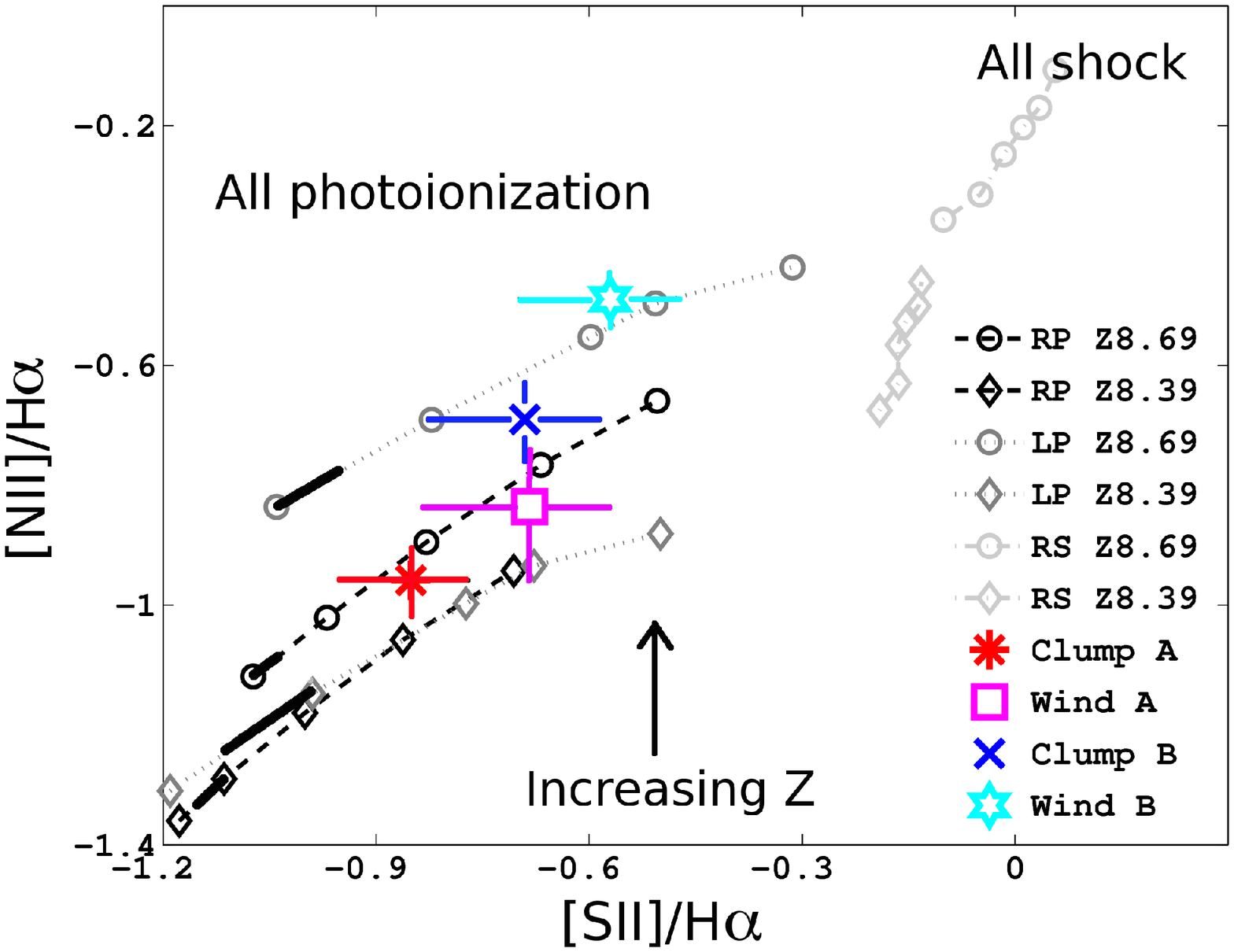}}
\caption{The \niiha vs. \siiha plane with data from the clump and wind regions as well as the photoionization and shock models (with \siiha calculated as described in section 2.3). The data are shown as a red asterisk, a purple square, a blue `x' and a cyan star with error bars (for clump A, wind A, clump B and wind B, respectively). The \cite{Ric+11} and \cite{Lev+10} photoionization and shock models are shown as the black and grey lines. The darkest/dashed lines are the \cite{Ric+11} photoionization models (RP), the medium/dotted lines are the \cite{Lev+10} photoionization models (LP) and the lightest/dash-dot lines are the \cite{Ric+11} shock models (RS). Open circles denote the models with O/H = 8.69 and open diamonds denote models with O/H = 8.39. For the photoionization models the points along the line represent a varying ionization parameter, increasing to the left (logU = 7--8). For the shock-only models, the points along the line vary in shock velocity, increasing upward (v = 100--200 \kms). The solid black lines indicate the ionization parameter range for each metallicity line as calculated in section 3.3. These measured ionization parameters are offset from those implied by the photoionization-only models, indicating a shock contribution to the emission.}
\end{figure}

We calculate the ionization parameter for the clumps using two different methods: 1) Using the O$_{32}$ diagnostic with metallicity-dependent fitting curves based on \cite{KewDop02} and 2) using the observed \ha luminosity and calculated electron density as follows:

\begin{equation}
\label{USFR}
U = \frac{Q}{4\pi r^{2}_{0}n_{e}}
\end{equation}

where Q = L$_{Lyc}$/h$\nu_{Lyc}$ is the rate of ionizing photons, L$_{Lyc}$ = 16xL$_{H\alpha}$ \citep[as given by][]{Ost89}, \ro is taken as the clump HWHM, and \nel is the electron density calculated from the \sii$~$ doublet. The results are shown in Table 2 and the range of ionization parameters (for the corresponding metallicities), is denoted by the thick black lines in Figure 7. These ionization parameters are 3-10 times higher than those predicted from the photoionization-only models of \cite{Ric+11} and \cite{Lev+10}. The data points appear to be offset towards the shock-only models, and are perhaps better fit by a model with contributions to the emission from both photoionization and shocks. In fact, if we compare the \niiha and \siiha ratios for the clumps along with the corresponding ionization parameters with Figure 10 from \cite{Ric+11}, which shows the photoionization + shock models, we find that the clumps and their winds are best fit by models with some shock contribution. The clumps are well fit by models with 5 $\pm$ 5\% and 10$^{+10}_{-5}$\% shock contribution to the nebular emission (for A and B), and the wind regions are well fit by models with 15$^{+25}_{-15}$\% and 30$^{+30}_{-20}$\% shocks, where the error bars come from fitting the 1$\sigma$ values of the ionization parameter and emission line ratios to the models. 

Based on the offsets from the observed ionization parameters, it is reasonable to assume that the elevated emission line ratios in the wind regions compared to the clumps are due to a larger contribution to the flux by shocks, and that the offset of the clumps in the BPT diagram from normal SFGs is due in part to shocks generated by star formation feedback. The highly turbulent, fast outflows emanating from the clumps could easily produce shocks as the hot superwind fluid collides and mixes with cooler ISM gas. 

\section{Discussion}
\subsection{Properties of the clumps and their outflows} 

Based on the observed properties of the clump A and B winds, we can perform simple calculations to better understand the structure of these outflows and their relationship to local and high-z galactic-scale superwinds. For the following calculations, we use the SFRs, gas masses and mass outflow rates as calculated in \cite{Gen+11}. The SFR was calculated from the \ha flux using the \cite{Ken98} calibration adjusted for a Chabrier IMF with the extinction correction described in Section 3.1. 

The gas masses were derived from the SFRs using the Schmidt-Kennicutt relation as given in \cite{Gen+10}. There are systematic errors that arise from this calculation due to (1) scatter in the intrinsic relation, (2) the varying slope and zero point of the relation as quoted by different authors, (3) the extrapolation of the relation from low-z to high-z, and (4) the application of the relation on $\sim$kpc scales. We address each of these issues in Appendix A, and ultimately derive an overall systematic uncertainty for the gas masses of the clumps of +0.3/-0.8 dex.

The mass outflow rates were calculated assuming an outflow of warm ionized gas into solid angle $\Omega$ with radially constant outflow rate and velocity as derived in \cite{Gen+11} and are quoted in Table 3. Details of the model are discussed in Appendix B. We note that for our outflow model, we only calculate the mass and mass outflow rate of the warm ionized component of the wind, and thus our numbers are lower limits.

\subsubsection{The Winds from Individual Clumps Are Very Massive and Energetic}

From our mass outflow rate derivation, we also obtain the mass of the warm ionized component in the wind (\Mw) for each clump. We estimate the energy of the wind as  \Ew = $\frac{1}{2}$ \Mw x \vout$^{2}$, where \vout$~$ is the maximum outflow velocity determined from the width of the \ha emission line (see Table 3). Using an average wind velocity instead would only change \Ew$~$ by a factor of 2 and will not affect any of the forthcoming conclusions. 

We find that the mass and energy in the outflows of individual clumps lies in the range 1--2x10$^{8}$ \msun$~$ and 3--7x10$^{56}$ ergs, several times greater than what is observed in the warm component from entire ULIRG and dwarf starburst galaxies (\Mw $\sim$ 10$^{5-7}$ \msun$~$ and \Ew $\sim$ 10$^{53-55}$ ergs) based on numbers compiled by \cite{Vei+05}. We caution that our measurements must be carefully compared to those of local starbursts, as the mass estimates are strongly dependent on the extinction corrections used and the quoted local starburst energy estimates are taken from dwarf starbursts and spiral galaxies, which likely produce less energetic winds than ULIRGs. 

The main point that should be gleaned from this section is that these clump-scale outflows are massive and very energetic even by local starburst standards. 

\subsubsection{Are the Outflows Energy or Momentum Driven?}

Galactic-scale outflows are considered to be driven either by the energy released from supernovae and the UV luminosity from stars \citep{McKeeOst77,Mat02} or through momentum transport by radiation pressure on dust and the momentum from stellar winds and supernovae \citep{Mur+05,Mur+10}. Momentum-driven winds tend to dominate in dense regions where much of the energy is radiated away, but the momentum remains \citep{Mur+05}. \cite{Mur+05} make several predictions for what one would expect given energy- or momentum-driven outflows from star-formation feedback in terms of \mdotout (the mass outflow rate), \pdotout (the momentum outflow rate) and $\eta$ (the mass loading).

For a starburst in which 90\% of the supernovae (SN) energy is radiated away, \cite{Mur+05} predict that $\dot{\rm E}_{\rm SN,out}\sim$ 10$^{-3}$L$_{bol}$, as 1\% of the total luminosity released throughout a starÕs lifetime is ejected in the supernova, and 10\% of the supernova energy is not radiated away. Thus, assuming the supernovae energy goes into the wind, $\dot{\rm E}_{\rm wind}$/L$_{bol} \sim$ 10$^{-3}$. For the clumps, we find that\\

$\dot{\rm E}_{\rm wind}$/L$_{bol} \sim \frac{1}{2}$ \mdotout  x \vout$^{2}$ / L$_{bol} = $ 0.005 and 0.05 with ranges of 7.6x10$^{-5}$--0.05 and 0.0002--0.3 for clumps A and B respectively.\\\\
Thus for a feedback efficiency of 10\%, the clump A wind could just barely be explained by an energy-driven model. However, the clump B wind is inconsistent with a purely energy-driven wind unless none of the SN energy is radiated away. If however, the extinction in clump B is much greater than for clump A, L$_{bol}$ from the disk would be underestimated more than \mdotout, which is coming from gas that is farther from the disk. This could lower $\dot{\rm E}_{\rm wind}$/L$_{bol}$ for clump B, bringing it closer to a number that could be explained by energy-driven winds. In addition, the large error bars for these quantities (mostly derived from the large error in \mdotout) do make them roughly consistent with the model, although due to the fact that \mdotout$~$ is likely an underestimate (see Section 4.1), the results are more likely skewed to larger values of $\dot{\rm E}_{\rm wind}$/L$_{bol}$.

For a purely momentum-driven wind, \cite{Mur+05} predict that $\dot{\rm p}_{\rm wind}$/P$_{rad} \sim$ 1, and we find that\\

$\dot{\rm p}_{\rm wind}$/P$_{rad} \sim$ \mdotout x \vout / (L$_{bol}$/c) = 6 and 30 with ranges of 0.5--65 and 2--245 for clumps A and B respectively.\\\\
This result indicates that clump A can also be explained by a momentum-driven wind, but clump B cannot ($\dot{\rm p}_{\rm wind}$/P$_{rad}$ = 30), although again, the large error bars make this conclusion somewhat uncertain.

There are many factors that could be biasing this measurement. First, as mentioned in \cite{Gen+11}, our outflow velocities could be too high by a factor of 2, depending on the definition chosen for the maximum outflow velocity.  Second, in an optically thick medium, the total effect of radiation pressure could be greater by a factor of (1+$\tau_{IR}$) due to multiple scattering of infrared photons, although it will likely not exceed a factor of a few. These two effects will bring down the value of $\dot{\rm p}_{\rm wind}$/P$_{rad}$. On the other hand, our mass outflow rates could be underestimated, both as a result of only including the warm ionized component of the wind and also due to overestimated wind densities (see next section), and this would mean our momentum wind efficiency is too low. These uncertainties are represented in the large systematic errors for $\dot{\rm p}_{\rm wind}$/P$_{rad}$. 

Our results suggest that the clump A wind could be either energy- or momentum-driven but that the clump B wind is harder to explain by either driver. However, it is possible that current models do not fully encapsulate the physics in these extreme high-z star-forming regions. In the next section we further compare the relative contributions to feedback by SN and radiation pressure using the \cite{OstShe11} model. We note that \cite{Mur+05} predict that in both energy and momentum driven outflow models, the mass loading factor ($\eta$ = \mdotout/SFR) $\sim$ 1 and this is roughly consistent with our observations ($\eta$ = 3--8). Similarly, \cite{Hop+11b} predict mass-loading factors in star-formation driven winds of 0.5--2. We compare these results to those of other high-z galaxies and local starbursts in Section 4.3.

\subsubsection{The Source of the High Velocity Dispersions}

We also investigate the pressure balance in the star-forming clumps to determine whether star-formation feedback is a viable source of the large velocity dispersions seen throughout the galaxy and by extension, in other high-z galaxies. We calculate the contributions to the gas pressure from the weight of the gas at the midplane (\Pw), from star-formation generated turbulence (\Pturb), from radiation pressure (\Prad) and from the observed velocity dispersion (\Pkin) from the model of \cite{OstShe11}, hereafter OS11. The model of OS11 is based on the assumption that in equilibrium, the weight of the gas balances the pressure from feedback. In the model, there are two different regimes defined by the surface density, where the feedback is either dominated by turbulence or radiation pressure; turbulence dominates in the galactic center from \siggas $\sim$ 100--10$^{4}$ \msunpc, and radiation pressure takes over for \siggas \textgreater 10$^{4}$ \msunpc. The pressures are defined as follows:

\begin{equation}
\label{pkin}
P_{kin} = \rho_{0} \sigma^{2}_{z}
\end{equation}
\begin{equation}
\label{pweight}
P_{weight} = \pi G \Sigma^{2} / 2 
\end{equation}
\begin{equation}
\label{pturb}
P_{turb} = \frac{f_{p}}{4} \frac{p_{*}}{m_{*}} \Sigma_{SFR} 
\end{equation}
\begin{equation}
\label{prad}
P_{rad} = \epsilon c \kappa_{IR} \Sigma \Sigma_{SFR} / 4 \\
\end{equation}
where $\rho_{0}$ is the midplane density (which we take as the electron density), $\sigma_{z}$ is the observed (narrow-line) velocity dispersion, $\Sigma$ and $\Sigma_{SFR}$ are the surface densities of gas and the SFR, f$_{p}$ is a factor of order 1-2 which describes the strength of turbulent dissipation, p$_{*}$/m$_{*}$ is the mean momentum injected by SN per unit mass of stars formed (which we take as 3000 \kms, following OS11), $\epsilon$ is the mass-to-radiation conversion efficiency from stars (which we take as 6.2x10$^{-4}$, following OS11) and $\kappa_{IR}$ is the mean opacity (which we take as 10 cm$^{2}$g$^{-1}$, following OS11).

From these calculations, we find that for clump A,  \Pkin$~$ balances \Pw, as proposed by OS11, and that \Pturb $\sim$ \Prad $~$ (see Table 3 for details). This latter finding is consistent with the fact that clump A is close to the boundary between the turbulent and radiation pressure dominated regimes defined by OS11, with $\Sigma \sim$ 8000 \msunpc. However, our results are inconsistent with the OS11 model in that \Pturb + \Prad \textless \Pkin $~$ (by about an order of magnitude, even with the errors), meaning that the pressure generated from star-formation feedback, both from SN and radiation pressure, does not fully account for the observed turbulence in the disk. This result also follows from applying equation 22 of OS11 to the star-formation efficiency derived from \cite{Gen+10} of 0.017, wherein the OS11 model predicts an observed velocity dispersion of 19--37 \kms (where the range reflects the the choice of f$_{p}$), while we observe dispersions of 50--100 \kms.

This issue of the source of the high dispersions in high-z galaxies has been explored by several authors \citep{Gen+08,Gen+11,Gre+10,Law+09,Epi+09,Kas+07,Bur+10} with no consensus yet reached concerning the dominant mechanism. Our conclusions on this matter are dependent on the exact choice of parameters used, on the uncertainty in our observations, and on the method applied to derive the intrinsic dispersion \citep{Dav+11}. However, we are fairly confident in our calculation of \Pkin, as the midplane density that we derive from the \sii$~$ doublet is roughly consistent with what is observed in dense star-forming regions in other galaxies \citep{Kew+01a}, and several authors have found that the high velocity dispersions observed from \ha in z $\sim$ 1-2 galaxies is consistent with the dispersions in molecular gas as observed from CO; for example \cite{Tac+10}; Tacconi et al. in prep for normal SF-galaxies and \cite{Tac+08,Engel+10} for SMGs.

For clump B, we find slightly different results. As for clump A, \Pturb + \Prad \textless \Pkin, but unlike for clump A, \Pkin \textgreater \Pw$~$ and \Pturb \textgreater \Prad. We note that the first two relations are less significant than for clump A since the errors are much higher as a result of the more uncertain electron density measurement. The fact that \Pturb \textgreater \Prad$~$ can be explained by the lower surface density of clump B (1200 \msunpc) which falls squarely in the turbulence-dominated regime (100--10$^{4}$ \msunpc). The results from these calculations support the claim of OS11 that the weight of the gas balances the turbulence in the disk, which suggests a self-regulatory mechanism for SF feedback. However, we find that while SF feedback does contribute to the observed turbulence it cannot fully account for it. 

We can look at this question in a different way, by comparing the energy injected into the wind through shocks with the turbulent energy in the clumps. To this end, we estimate the energy from shocks:

\begin{equation}
\label{eshock}
E_{shock} = E_{w}  f_{shock},\\
\end{equation}
where \Ew$~$ is given above and f$_{shock}$ is the fraction of this radiated energy that comes from shocks. Based on our analysis in Section 3.3, we choose f$_{shock}$ values of 0.2 and 0.3 for clumps A and B, respectively. The turbulent energy in the disk is:

\begin{equation}
\label{eturb}
E_{turb} = \frac{3}{2} M_{gas}  \sigma^{2}_{z}\\
\end{equation}
where M$_{gas}$ is the clump gas mass. We find for the \Ew$~$ estimate (\Ew = 1/2 \Mw x v$^{2}_{out}$) that E$_{shock}$/E$_{turb}$ is 0.011$^{+0.12}_{-0.014}$ for clump A and 0.11$^{+0.88}_{-0.23}$ for clump B. This suggests that the energy from shocks cannot fully account for the energy in turbulence, but in the case of clump B, it can contribute significantly. 

This conclusion is consistent with that above, wherein SF feedback is an important component to the large velocity dispersions, but another strong component is necessary. Even the sum of the contributions of star-formation feedback (as given by OS11) and shocks falls short of the observed energy from turbulence. However, if the density of the wind was overestimated as suggested in the next section, then the corrected mass and energy in the outflow would be $\sim$ 25 times larger, bringing E$_{shock}$/E$_{turb} \sim$ 1 (as represented by the upper error bar for clump B). In addition, the E$_{shock}$ estimate only includes the energy from the warm ionized component in the wind, and the hot, neutral and molecular components could also contribute to E$_{shock}$. Therefore, future observations that constrain the contribution to the wind from these other components is critical to understanding energy balance. 

Assuming our density estimates and mass outflow rates are correct, the combination of feedback and shocks cannot fully explain the high velocity dispersions observed in high-z galaxies. Roughly 85--90\% of the velocity dispersion remains unaccounted for and must have some other source, such as clump-clump interaction, clump-stream interaction, or the release of gravitational energy as gas or clumps fall inward.

\subsubsection{The Outflow Can Escape the Disk of the Galaxy}

To test whether or not the outflow can escape the plane of the galaxy, we compare the pressures derived from the OS11 model with the ram pressure in the wind:

\begin{equation}
\label{pram}
P_{ram} = \rho_{w} v^{2}_{out}\\
\end{equation}
where $\rho_{w}$ is the density in the wind and is somewhat uncertain. \cite{Gen+11} adopted a value of 100 cm$^{-3}$ based on \sii$~$ doublet observations of local outflows from \cite{HecArmMil90} and \cite{LehHec96}. However, more recent work has shown that the actual density of the warm ionized component in superwinds could be almost 2 orders of magnitude lower \citep{Cec+01}. Using the larger density estimate (100 cm$^{-3}$) for clump A, P$_{ram} \sim$  \Pw, which suggests that the wind would barely be able to break out of the plane of the disk, while for the lower density estimate (4 cm$^{-3}$), P$_{ram} << $ \Pw. The situation is reversed for clump B which has a more evolved and therefore faster wind (see models by \cite{HecArmMil90} which suggest that winds become faster with increasing radius), such that both density estimates yield P$_{ram} >> $ \Pw. In this case, the wind would have no trouble escaping the disk. 

Alternatively, we can estimate the wind density by assuming that P$_{ram}$ balances \Pw. With the pressures calculated as before, we get n$_{wind} $ = 67 $\pm$ 74 cm$^{-3}$ for clump A and 0.4 $\pm$ 0.5 cm$^{-3}$ for clump B. This result is consistent with clump B having a more extended and therefore more diffuse outflow.

We can also test whether or not gas can escape the galaxy by comparing the outflow velocities of the clump winds with the escape velocities of the clumps and the galaxy. At the distance of the clump HWHM and estimating the clump mass as 2 x M$_{gas}$ (assuming half of the clump mass is in stars), we find that for clumps A and B \vout/v$_{esc}$ = 0.75 $\pm$ 0.39 and 2.4 $\pm$ 1.3, respectively. Therefore, some of the warm ionized gas could escape the gravitational potential of the clumps. In fact, some of this material could escape the galaxy as well since \vout/v$_{esc,galaxy}$ = 0.54 $\pm$ 0.26 and 0.99 $\pm$ 0.48 for clumps A and B. Thus analysis of both the ram pressure of the wind and \vout/v$_{escape}$ suggests that a significant fraction of the gas could escape the gravitational potential of the clumps distributing the warm ionized gas elsewhere in the galaxy and even into the IGM.

\subsubsection{How Clumpy are the Clumps?}

We can estimate the `clumpiness' of the clumps (or a volume filling factor of high density gas) using the observed electron density (from the \sii$~$ doublet) and the calculated surface density. This number can be compared to that for Milky Way GMCs and central starburst regions to better understand how the star-forming regions are distributed within the clumps. We define it as follows:

\begin{equation}
\label{clumpiness}
clumpiness = \frac{n_{avg}}{n_{local}} = \frac{\Sigma}{4/3 r_{cl} \mu m_{h}}  \frac{1}{n_{H_{2}}}
\end{equation}

\begin{equation}
n_{H_{2}} = n_{HII}  \frac{T_{HII}}{T_{H_{2}}}\\
\end{equation}
where the latter equation is simple thermal pressure balance to obtain the density in the colder molecular gas from the hotter ionized gas, and n$_{HII}$ is the local electron density derived from the \sii$~$ doublet. We note that this thermal pressure balance gives us an upper limit for the molecular density since it neglects any non-thermal pressure due to turbulent cloudlet motion, which would likely be more similar for the ionized and molecular phases. A volume filling (clumpiness) factor of 1 would mean that the high density star-forming gas occupies all of the volume of the clumps, whereas a factor of 0 would mean it takes up none of the volume. We set T$_{HII} $ = 10$^{4}$ K and T$_{H_{2}} $ = 30 K \citep{Elb+11,Hwa+11} to obtain n$_{H_{2}} $ = 6x10$^{5}$ and 1x10$^{5}$ cm$^{-3}$ for the clumps. For clumps A and B, we get volume filling factors of 3.7x10$^{-4}$ and 2.3x10$^{-4}$ with 1$\sigma$ ranges of 0--7.6x10$^{-4}$ and 0--6.3x10$^{-4}$, respectively. If, however, we adopt the molecular density found by \cite{Dan+11} for a lensed z $\sim$ 2.3 SMG of 10$^{4}$ cm$^{-3}$, we find much higher volume filling factors of 0.022 and 0.0023 for clumps A and B, respectively. These different values serve as our lower and upper limits for the volume filling factors of the clumps and are reflected in the errors (Table 3).

The volume filling factors based on the local electron density of the clumps are almost 2 orders of magnitude smaller than what we find in Milky Way (MW) GMCs. Assuming an average density $\sim$ 100 cm$^{-3}$ and a local/core density $\sim$ 10$^{4}$ \citep{StaPal05}, then n$_{avg}$/n$_{local}$ (MW) $\sim$ 10$^{-2}$. Our clump estimates, however, are more similar to the volume filling factor in local starburst galaxies, based on models of M82. In the cloud-cluster starburst region model of \cite{For+01}, they estimate that clouds of molecular material are roughly 0.5 pc in radius surrounded by a layer of neutral and ionized gas up to 1 pc with a spacing of 2--7 pc between them. This gives an average volume filling factor of 3.6x10$^{-3}$ for the cold molecular gas. This suggests that the star forming regions in the clumps are more concentrated than regions in both MW GMCs and starburst regions unless we adopt the \cite{Dan+11} molecular gas density, in which case, our values are comparable to those from MW GMCs and local starbursts.

This should not be surprising as the molecular densities in the clumps are over an order of magnitude larger than what is found in star-forming cores in MW GMCs and the pressure in the disk (as seen in 4.1.3 and Table 3) is several orders of magnitude larger than what is found in the MW (P $\sim$ 10$^{-10}$ ergs cm$^{-1}$) and is more comparable to the pressure in local ULIRGs (P $\sim$ 10$^{-7}$ ergs cm$^{-1}$) \citep{DowSol98}. Thus, the clumps more closely resemble starburst regions in ULIRGs than MW GMCs, both in terms of physical properties and spatial distribution of dense gas. This result implies that the very large SFRs of the clumps are not because a higher fraction of the volume is involved in star-formation (in fact less volume is forming stars than for MW GMCs), but because the regions that are forming stars have much higher gas fractions.

\subsection{Do the Clumps Represent an Evolutionary Sequence?}

In this section, we explore the evolution of the clumps through their winds and whether they could typify different stages in clump development and evolution. We have seen throughout this paper that while both clumps A and B are actively forming stars and produce powerful winds, their individual properties are quite different. Clump B contains more enhanced shock-like emission line ratios and broad emission spatially offset (and therefore further away) from the clump as well as larger line widths, both indicative of a more mature outflow. The clumps also differ in their volume and surface densities (clump A is denser). This could be the result of the expulsion of high density gas from star formation feedback from clump B or it could be due to the fact that the luminosity and density measurements from clump B include not only clump gas but a large amount of the more diffuse gas from the wind. Both explanations further support the presence in clump B of a more mature outflow, suggesting that clump B has been forming stars for a longer period of time and thus clump B is older than clump A. 

This notion is supported by the difference in gas phase metallicity between the clumps as calculated from \niiha which suggests an age difference. We calculate how long it would take clump A to be as metal enriched as clump B (assuming the \niiha ratio reflects the abundance only and is unaffected by shocks as a simplification). Using the closed and leaky box models outlined in \cite{Erb08} \citep[see][Appendix C]{Gen+11}, we find this timescale ranges from 170 Myr  -- 1.7 Gyr. If the value of \niiha is more strongly affected by shocks for clump B (as is suggested by Figure 7), then the resulting timescale would be reduced, but it is likely that at least some of the \niiha enhancement in clump B is due to a higher metallicity, and thus more mature clump.

Given that clump B is likely older than clump A, we estimate how long it would take a clump A like object to turn into clump B, which has half of the clump A M$_{gas}$, assuming that the clumps start out with roughly the same M$_{gas}$. Although the gas masses are uncertain by 70--85\% due to the fact that the physical conditions in these $\sim$kpc size clumps could be very different from the galaxies from which we derived the KS relation, it is hard to imagine that the relative masses between the clumps would be that far off, given that the physics and CO to H conversion factor in both the clumps are likely similar. If all of the extra gas from clump A is turned into stars, this would take $\sim$ 200--700 Myr (with the range reflecting the SFRs of clumps A and B). However, if outflows are removing gas from the clumps as well, then this timescale is much shorter; with an $\eta$ of 1, the timescale is twice as fast and with an $\eta$ of 3 (corresponding to clump A), the timescale is four times as fast. In Section 4.1.4, we found that a significant fraction of the outflowing gas could be lost from the clumps and perhaps even the galaxy. Based on our finding that $\eta >$ 1, the age difference between the clumps obtained from the SFR and M$_{gas}$ is at most 350 Myr, and for this to be consistent with our results based on \nii/H$\alpha$, shocks must play an important role in the emission.

We can also compare clumps A and B with C. Clump C has a similar flux and SFR as clump B, and the extended \nii$~$ emission from clump B continues to the region around clump C, indicating that it is somewhat enriched from star-formation. However, clump C has a narrow \ha line profile compared to clumps A and B and has no indication of outflows. Perhaps clump C could be placed further down the evolutionary sequence of clumps A and B: it is declining from itÕs major episode of star formation and its outflows are too diffuse to be observed or are projected onto clump B due to the inclination of the galaxy.

\subsection{Comparison of Star-forming and Outflow Properties of ZC406690 to those of Galaxies at Low- and High-z}

High-z SFGs, like ZC406690, in many ways resemble both local starburst galaxies and local normal disk galaxies. The results from this work and others suggest that the overall structure and star-forming history (steady as opposed to bursty) of these galaxies more closely resembles normal local SFGs \citep{Wuy+11b,Elb+11}, while properties of the star-forming regions and their outflows are more comparable to those of local nuclear starbursts.

The overall structure and large-scale properties of high-z rotation-dominated SFGs are in many ways similar to local disk galaxies. First, they both have disk-like morphologies (although local galaxies have much thinner disks) and most of the star-formation in these high-z galaxies occurs in an extended disk or ring and not in a central and compact starburst region. Second, they are not considered extreme star-formers for their redshift (unlike SMGs), and many of these observed galaxies have kinematic structures suggestive of no major interactions \citep{Gen+06,Gen+08,Gen+11,Sha+08,For+09,Cre+09,Gne+10,Jon+10}. Third, with the steady accretion of cold gas (in agreement with the \lcdm$~$ model), these high-z galaxies can sustain star formation for long timescales ($\sim$ 1-2 Gyr).

In local properties, however, high-z SFGs more closely resemble local starbursts. From the BPT diagram (Figure 6, as well as \cite{Shapley+05,Erb+06,Liu+08}), we see that normal SFGs at z$\sim$1--3 are offset from normal local SFGs and better coincide with the region occupied by local starburst galaxies. As shown by \cite{Liu+08} and \cite{Bri+08}, this is due in part to variations in ionization parameter, SFR, sSFR, and density, between normal SFGs at high and low-z, and that in terms of these parameters local starbursts are more similar to the high-z SFGs. It should be noted that in the \cite{Liu+08} and \cite{Bri+08} studies, they compared SDSS galaxies that were part of the main star-forming branch and ones that were offset from this branch, using similar stellar mass ranges for both samples.  If we compare the properties (SFR, \siggas, P, n$_{H_{2}}$, $\sigma$, U, volume filling factor) of clumps A and B of ZC406690 with local galaxies, we find a closer correspondence with starbursts than with normal star-forming galaxies (see Table 4). 

The outflows are dictated by the properties of the star-forming regions and not the overall structure of the galaxy, and are therefore similar to winds observed from local starbursts. We find that the mass-loading factor for the clumps is consistent with those observed from local starburst galaxies: 0.26 for a sample of several IRGs and ULIRGs \citep{RupVeiSan05} and 2.5 for Mrk 231 \citep{RupVei11}. In addition, for Mrk 231, $\dot{\rm E}_{\rm wind}$/L$_{bol} \sim$ 0.007 \citep{RupVei11}, which is similar to our estimate for clump A. The measurements for Mrk 231 are lower limits; thus if the mass-loading and $\dot{\rm E}_{\rm wind}$/L$_{bol}$ factors are larger, then Mrk 231 would be more similar to clump B rather than clump A. It is important to note that the measurements for Mrk 231 may not be directly comparable to the observations presented here, as they are made using the NaID absorption feature and are therefore sampling a different (and possibly more massive and energetic) component of the outflow. We also find similar mass loading factors when comparing our observations ($\eta \sim$ 3--8) to those of warm ionized outflows from local dwarf starbursts, where \cite{Mar99} finds $\eta \sim$ 1--5.

The outflow properties of individual clumps in ZC406690 resemble those from integrated data of other high-z SFGs. As for the local starbursts, most outflows from high-z galaxies have $\eta \ge$ 1 \citep{Pet+00,Wei+09,Ste+10}. We also find that the clumps fit the \vout-M$_{*}$ and \vout-SFR relations found at z$\sim$1.4 by \cite{Wei+09}. In addition, in their study of high-z outflows observed in absorption of light from background galaxies, \cite{Ste+10} find maximum outflow velocities of 700--800 \kms, similar to what is found for the more rapid clump B outflow (810 \kms). These similarities suggest that the outflows observed in previous work from galaxy-integrated spectra could originate from massive clumps in those galaxies. They also indicate that the star-forming properties in high-z galaxies could scale from clump to galaxy size, suggesting that the lessons learned from studying individual star-forming clumps could be applied to entire galaxies and vice versa. 

If some of the gas can escape the gravitational potential of these galaxies (as is likely for the hot and fast wind fluid, see \cite{StrHec09} and perhaps also part of the warm ionized component), these numerous and powerful winds are likely significant in enriching the IGM at early times. Such enrichment from outflows is supported by theoretical models; \cite{Erb08} found that in their models of metal enrichment by star-formation paired with infall and outflow, $\eta \sim$ 1 is required to fit the mass-metallicity relation at z $\sim$ 2. In addition, the results of cosmological simulations by \cite{Dave+11} indicate that $\eta \sim$ 1 is a necessary condition to reproduce the mass-metallicity relation as well as IGM abundances. On the observational side, \cite{Ste+10} find enriched gas as far as 125 kpc away from nearby galaxies in a sample of $\sim$100 z$\sim$2--3 Lyman-break SFGs. On the other hand, if some of the gas falls back onto the clumps (as suggested by the fact that v$_{wind} <$ v$_{esc,clump}$ for clump A), this could prolong the clump star-formation timescales.

Evidence for shocks in the clump outflows allows us to draw additional comparisons between ZC406690 and local and high-z SFGs. The finding that some shock contribution to the emission is necessary to explain the observed emission line ratios in the clumps and their winds is consistent with observations of outflows from local starbursts, which are dominated by shock-like emission line ratios even in the absence of AGN. However, if shocks are a common feature of high-z outflows, which in turn are ubiquitous in z$\sim$2--3 galaxies, then this has important consequences for abundance measurements at high-z. As suggested by \cite{Shapley+05} and \cite{Erb+06}, using \niiha$~$ alone as an abundance indicator could be flawed in the presence of shocks or AGN. As they note, many of these emission line ratio abundance indicators are calibrated from galaxies and star-forming regions that populate the ÔHIIÕ branch of the BPT diagram. Therefore, galaxies that are offset from this branch (mainly in terms of \niiha), such as local starbursts and high-z SFGs, require a different calibration. Another option is to use photoionization and shock models with multiple emission line diagnostics to get more accurate abundances. It is not yet clear whether abundances are overestimated due to shocks in most observed high-z SFGs, but this is an issue that deems more attention through continued high-resolution AO emission line measurements in these systems.

\section{Conclusions}

Using high S/N SINFONI/VLT observations with AO, we find powerful outflows localized to the star-forming $\sim$kpc-size clumps of the \ztwo SFG, ZC406690. Spectra of the clumps and surrounding regions have broad, blue \ha line wings extending up to 800 \kms from line center. Much of this broad emission is coming from a region offset from the center of clump B that also has enhanced \niiha and \siiha ratios, indicative of a superwind laden with shocks. The outflow velocities derived from the line widths are comparable to and slightly in excess of the escape velocity of the clumps (and the galaxy), indicating that some of the wind material could fall back on other parts of the galaxy or perhaps escape out to the IGM.

By comparing the energy and momentum injection rates of the wind to the \cite{Mur+05} model of a starburst-driven wind, we find that the clump A wind satisfies the predictions for both energy- and momentum-driven winds, but that the clump B wind cannot be explained by either within the context of this model. Star formation feedback from the winds could greatly contribute to the large velocity dispersions seen in ZC406690 and other z $\sim$ 2--3 galaxies, although based on measurements from the clumps, feedback alone cannot account for this turbulence. 

The clumps have a large range in ages (from metallicity), which is reflected in the maturity of their outflows (as seen in the variation in outflow velocity, physical extent, mass-loading and shock-like emission line ratios), and they appear to be more extincted than the surrounding regions, as one would expect for dense star-forming regions. It is possible that the younger clump (A) resembles what clump B looked like at an earlier time and that the current state of the galaxy gives us a snapshot of various stages of clump formation and evolution.

In terms of global properties (disk-like morphology, location of star-formation, star-formation timescale), ZC406690 is similar to local disk galaxies, while in terms of local properties of the star-forming regions and the outflows (SFR, $\Sigma_{SFR}$, n$_{H_{2}}$, pressure, ionization parameter, \vout, mass-loading, volume filling factor), its clumps more closely resemble star-forming regions in local starburst galaxies. The outflow properties of the clumps are also similar to those of outflows observed in other galaxies at z $\sim$ 1--3, suggesting that massive clumps could be the source of a significant fraction of outflowing material in these galaxies as well, and that although ZC406690 is quite massive, it may not be very unusual for \ztwo SFGs in terms of star-formation feedback. 

These types of winds coming from kpc-size regions could have a significant effect on galaxy evolution during this epoch. They could inhibit star-formation with the injection of heat into the ISM and the mass-loss associated with the outflowing material in the wind could reduce the lifetime of the clumps. They also influence the chemical composition of the ISM and the IGM by expelling metal-rich gas from star-forming regions. The results presented in this paper inform our understanding of the properties, evolution, and feedback of star-forming regions at high-z, which in turn teaches us about the role that these star-forming regions play in galaxy evolution on Gyr timescales.  

\acknowledgements
We thank the ESO staff, especially those at Paranal Observatory, for their ongoing support during the many past and continuing observing runs over which the SINS project is being carried out. We also acknowledge the SINFONI and PARSEC teams, whose devoted work on the instrument and laser paved the way for the success of the SINS observations. We would also like to thank Mike Dopita and Emily Levesque for sharing the data from their photoionization and shock models and for useful discussions regarding those models. We would also like to thank Sylvain Veilleux, Amiel Sternberg and Eliot Quataert for helpful conversations concerning this work. SFN is supported by an NSF grfp grant. CM, AR and DV acknowledge partial support by the ASI grant ``COFIS-Analisi Dati'' and by the INAF grant ``PRIN-2008''.


\begin{thebibliography}{120}
\expandafter\ifx\csname natexlab\endcsname\relax\def\natexlab#1{#1}\fi

\bibitem[{{Adelberger} {et~al.}(2005){Adelberger}, {Shapley}, {Steidel},
  {Pettini}, {Erb}, \& {Reddy}}]{Ade+05}
{Adelberger}, K.~L., {Shapley}, A.~E., {Steidel}, C.~C., {Pettini}, M., {Erb},
  D.~K., \& {Reddy}, N.~A. 2005, \apj, 629, 636

\bibitem[{{Adelberger} {et~al.}(2003){Adelberger}, {Steidel}, {Shapley}, \&
  {Pettini}}]{Ade+03}
{Adelberger}, K.~L., {Steidel}, C.~C., {Shapley}, A.~E., \& {Pettini}, M. 2003,
  \apj, 584, 45

\bibitem[{{Adelman-McCarthy} \& {et al.}(2009)}]{Ade+09}
{Adelman-McCarthy}, J.~K. \& {et al.} 2009, VizieR Online Data Catalog, 2294, 0

\bibitem[{{Baldwin} {et~al.}(1981){Baldwin}, {Phillips}, \&
  {Terlevich}}]{BPT81}
{Baldwin}, J.~A., {Phillips}, M.~M., \& {Terlevich}, R. 1981, \pasp, 93, 5

\bibitem[{{Bigiel} {et~al.}(2011){Bigiel}, {Leroy}, {Walter}, {Brinks}, {de
  Blok}, {Kramer}, {Rix}, {Schruba}, {Schuster}, {Usero}, \&
  {Wiesemeyer}}]{Big+11}
{Bigiel}, F., {Leroy}, A.~K., {Walter}, F., {Brinks}, E., {de Blok}, W.~J.~G.,
  {Kramer}, C., {Rix}, H.~W., {Schruba}, A., {Schuster}, K.-F., {Usero}, A., \&
  {Wiesemeyer}, H.~W. 2011, \apjl, 730, L13

\bibitem[{{Bolatto} {et~al.}(2008){Bolatto}, {Leroy}, {Rosolowsky}, {Walter},
  \& {Blitz}}]{Bol+08}
{Bolatto}, A.~D., {Leroy}, A.~K., {Rosolowsky}, E., {Walter}, F., \& {Blitz},
  L. 2008, \apj, 686, 948

\bibitem[{{Bonnet} {et~al.}(2004)}]{Bon+04}
{Bonnet}, H. {et~al.} 2004, in Presented at the Society of Photo-Optical
  Instrumentation Engineers (SPIE) Conference, Vol. 5490, Advancements in
  Adaptive Optics. Edited by Domenico B. Calia, Brent L. Ellerbroek, and
  Roberto Ragazzoni. Proceedings of the SPIE, Volume 5490, pp. 130-138 (2004).,
  ed. D.~{Bonaccini Calia}, B.~L. {Ellerbroek}, \& R.~{Ragazzoni}, 130--138

\bibitem[{{Brinchmann} {et~al.}(2008){Brinchmann}, {Pettini}, \&
  {Charlot}}]{Bri+08}
{Brinchmann}, J., {Pettini}, M., \& {Charlot}, S. 2008, \mnras, 385, 769

\bibitem[{{Burkert} {et~al.}(2010){Burkert}, {Genzel}, {Bouch{\'e}}, {Cresci},
  {Khochfar}, {Sommer-Larsen}, {Sternberg}, {Naab}, {F{\"o}rster Schreiber},
  {Tacconi}, {Shapiro}, {Hicks}, {Lutz}, {Davies}, {Buschkamp}, \&
  {Genel}}]{Bur+10}
{Burkert}, A., {Genzel}, R., {Bouch{\'e}}, N., {Cresci}, G., {Khochfar}, S.,
  {Sommer-Larsen}, J., {Sternberg}, A., {Naab}, T., {F{\"o}rster Schreiber},
  N., {Tacconi}, L., {Shapiro}, K., {Hicks}, E., {Lutz}, D., {Davies}, R.,
  {Buschkamp}, P., \& {Genel}, S. 2010, \apj, 725, 2324

\bibitem[{{Calzetti} {et~al.}(2000){Calzetti}, {Armus}, {Bohlin}, {Kinney},
  {Koornneef}, \& {Storchi-Bergmann}}]{Cal+00}
{Calzetti}, D., {Armus}, L., {Bohlin}, R.~C., {Kinney}, A.~L., {Koornneef}, J.,
  \& {Storchi-Bergmann}, T. 2000, \apj, 533, 682

\bibitem[{{Cecil} {et~al.}(2001){Cecil}, {Bland-Hawthorn}, {Veilleux}, \&
  {Filippenko}}]{Cec+01}
{Cecil}, G., {Bland-Hawthorn}, J., {Veilleux}, S., \& {Filippenko}, A.~V. 2001,
  \apj, 555, 338

\bibitem[{{Chabrier}(2003)}]{Cha03}
{Chabrier}, G. 2003, \pasp, 115, 763

\bibitem[{{Cole} {et~al.}(2000){Cole}, {Lacey}, {Baugh}, \& {Frenk}}]{Cole+00}
{Cole}, S., {Lacey}, C.~G., {Baugh}, C.~M., \& {Frenk}, C.~S. 2000, \mnras,
  319, 168

\bibitem[{{Cresci} {et~al.}(2009)}]{Cre+09}
{Cresci}, G. {et~al.} 2009, \apj, 697, 115

\bibitem[{{Daddi} {et~al.}(2010){Daddi}, {Elbaz}, {Walter}, {Bournaud},
  {Salmi}, {Carilli}, {Dannerbauer}, {Dickinson}, {Monaco}, \&
  {Riechers}}]{Dad+10}
{Daddi}, E., {Elbaz}, D., {Walter}, F., {Bournaud}, F., {Salmi}, F., {Carilli},
  C., {Dannerbauer}, H., {Dickinson}, M., {Monaco}, P., \& {Riechers}, D. 2010,
  \apjl, 714, L118

\bibitem[{{Daddi} {et~al.}(2004)}]{Dad+04}
{Daddi}, E. {et~al.} 2004, \apjl, 600, L127

\bibitem[{{Danielson} {et~al.}(2011){Danielson}, {Swinbank}, {Smail}, {Cox},
  {Edge}, {Weiss}, {Harris}, {Baker}, {De Breuck}, {Geach}, {Ivison}, {Krips},
  {Lundgren}, {Longmore}, {Neri}, \& {Flaquer}}]{Dan+11}
{Danielson}, A.~L.~R., {Swinbank}, A.~M., {Smail}, I., {Cox}, P., {Edge},
  A.~C., {Weiss}, A., {Harris}, A.~I., {Baker}, A.~J., {De Breuck}, C.,
  {Geach}, J.~E., {Ivison}, R.~J., {Krips}, M., {Lundgren}, A., {Longmore}, S.,
  {Neri}, R., \& {Flaquer}, B.~O. 2011, \mnras, 410, 1687

\bibitem[{{Dav{\'e}} {et~al.}(2011){Dav{\'e}}, {Finlator}, \&
  {Oppenheimer}}]{Dave+11}
{Dav{\'e}}, R., {Finlator}, K., \& {Oppenheimer}, B.~D. 2011, \mnras, 416, 1354

\bibitem[{{Davies} {et~al.}(2011){Davies}, {Forster Schreiber}, {Cresci},
  {Genzel}, {Bouche}, {Burkert}, {Buschkamp}, {Genel}, {Hicks}, {Kurk}, {Lutz},
  {Newman}, {Shapiro}, {Sternberg}, {Tacconi}, \& {Wuyts}}]{Dav+11}
{Davies}, R., {Forster Schreiber}, N.~M., {Cresci}, G., {Genzel}, R., {Bouche},
  N., {Burkert}, A., {Buschkamp}, P., {Genel}, S., {Hicks}, E., {Kurk}, J.,
  {Lutz}, D., {Newman}, S., {Shapiro}, K., {Sternberg}, A., {Tacconi}, L.~J.,
  \& {Wuyts}, S. 2011, ArXiv e-prints

\bibitem[{{Denicol{\'o}} {et~al.}(2002){Denicol{\'o}}, {Terlevich}, \&
  {Terlevich}}]{deNic+02}
{Denicol{\'o}}, G., {Terlevich}, R., \& {Terlevich}, E. 2002, \mnras, 330, 69

\bibitem[{{Devine} \& {Bally}(1999)}]{Dev+99}
{Devine}, D. \& {Bally}, J. 1999, \apj, 510, 197

\bibitem[{{Dopita} \& {Sutherland}(1995)}]{DopSut95}
{Dopita}, M.~A. \& {Sutherland}, R.~S. 1995, \apj, 455, 468

\bibitem[{{Downes} \& {Solomon}(1998)}]{DowSol98}
{Downes}, D. \& {Solomon}, P.~M. 1998, \apj, 507, 615

\bibitem[{{Eisenhauer} {et~al.}(2003)}]{Eis+03}
{Eisenhauer}, F. {et~al.} 2003, in Presented at the Society of Photo-Optical
  Instrumentation Engineers (SPIE) Conference, Vol. 4841, Instrument Design and
  Performance for Optical/Infrared Ground-based Telescopes. Edited by Iye,
  Masanori; Moorwood, Alan F. M. Proceedings of the SPIE, Volume 4841, pp.
  1548-1561 (2003)., ed. M.~{Iye} \& A.~F.~M. {Moorwood}, 1548--1561

\bibitem[{{Elbaz} {et~al.}(2011){Elbaz}, {Dickinson}, {Hwang},
  {D{\'{\i}}az-Santos}, {Magdis}, {Magnelli}, {Le Borgne}, {Galliano},
  {Pannella}, {Chanial}, {Armus}, {Charmandaris}, {Daddi}, {Aussel}, {Popesso},
  {Kartaltepe}, {Altieri}, {Valtchanov}, {Coia}, {Dannerbauer}, {Dasyra},
  {Leiton}, {Mazzarella}, {Alexander}, {Buat}, {Burgarella}, {Chary}, {Gilli},
  {Ivison}, {Juneau}, {Le Floc'h}, {Lutz}, {Morrison}, {Mullaney}, {Murphy},
  {Pope}, {Scott}, {Brodwin}, {Calzetti}, {Cesarsky}, {Charlot}, {Dole},
  {Eisenhardt}, {Ferguson}, {F{\"o}rster Schreiber}, {Frayer}, {Giavalisco},
  {Huynh}, {Koekemoer}, {Papovich}, {Reddy}, {Surace}, {Teplitz}, {Yun}, \&
  {Wilson}}]{Elb+11}
{Elbaz}, D., {Dickinson}, M., {Hwang}, H.~S., {D{\'{\i}}az-Santos}, T.,
  {Magdis}, G., {Magnelli}, B., {Le Borgne}, D., {Galliano}, F., {Pannella},
  M., {Chanial}, P., {Armus}, L., {Charmandaris}, V., {Daddi}, E., {Aussel},
  H., {Popesso}, P., {Kartaltepe}, J., {Altieri}, B., {Valtchanov}, I., {Coia},
  D., {Dannerbauer}, H., {Dasyra}, K., {Leiton}, R., {Mazzarella}, J.,
  {Alexander}, D.~M., {Buat}, V., {Burgarella}, D., {Chary}, R.-R., {Gilli},
  R., {Ivison}, R.~J., {Juneau}, S., {Le Floc'h}, E., {Lutz}, D., {Morrison},
  G.~E., {Mullaney}, J.~R., {Murphy}, E., {Pope}, A., {Scott}, D., {Brodwin},
  M., {Calzetti}, D., {Cesarsky}, C., {Charlot}, S., {Dole}, H., {Eisenhardt},
  P., {Ferguson}, H.~C., {F{\"o}rster Schreiber}, N., {Frayer}, D.,
  {Giavalisco}, M., {Huynh}, M., {Koekemoer}, A.~M., {Papovich}, C., {Reddy},
  N., {Surace}, C., {Teplitz}, H., {Yun}, M.~S., \& {Wilson}, G. 2011, \aap,
  533, A119+

\bibitem[{{Elmegreen} \& {Elmegreen}(2006)}]{ElmElm06}
{Elmegreen}, B.~G. \& {Elmegreen}, D.~M. 2006, \apj, 650, 644

\bibitem[{{Elmegreen} \& {Elmegreen}(2010)}]{ElmElm10}
---. 2010, \apj, 722, 1895

\bibitem[{{Elmegreen} {et~al.}(2009){Elmegreen}, {Elmegreen}, {Marcus},
  {Shahinyan}, {Yau}, \& {Petersen}}]{Elm+09a}
{Elmegreen}, D.~M., {Elmegreen}, B.~G., {Marcus}, M.~T., {Shahinyan}, K.,
  {Yau}, A., \& {Petersen}, M. 2009, \apj, 701, 306

\bibitem[{{Elmegreen} {et~al.}(2005){Elmegreen}, {Elmegreen}, {Rubin}, \&
  {Schaffer}}]{Elm+05}
{Elmegreen}, D.~M., {Elmegreen}, B.~G., {Rubin}, D.~S., \& {Schaffer}, M.~A.
  2005, \apj, 631, 85

\bibitem[{{Elmegreen} {et~al.}(2004){Elmegreen}, {Elmegreen}, \&
  {Sheets}}]{ElmElmShe04}
{Elmegreen}, D.~M., {Elmegreen}, B.~G., \& {Sheets}, C.~M. 2004, \apj, 603, 74

\bibitem[{{Engel} {et~al.}(2010){Engel}, {Tacconi}, {Davies}, {Neri}, {Smail},
  {Chapman}, {Genzel}, {Cox}, {Greve}, {Ivison}, {Blain}, {Bertoldi}, \&
  {Omont}}]{Engel+10}
{Engel}, H., {Tacconi}, L.~J., {Davies}, R.~I., {Neri}, R., {Smail}, I.,
  {Chapman}, S.~C., {Genzel}, R., {Cox}, P., {Greve}, T.~R., {Ivison}, R.~J.,
  {Blain}, A., {Bertoldi}, F., \& {Omont}, A. 2010, \apj, 724, 233

\bibitem[{{Epinat} {et~al.}(2009){Epinat}, {Contini}, {Le F{\`e}vre},
  {Vergani}, {Garilli}, {Amram}, {Queyrel}, {Tasca}, \& {Tresse}}]{Epi+09}
{Epinat}, B., {Contini}, T., {Le F{\`e}vre}, O., {Vergani}, D., {Garilli}, B.,
  {Amram}, P., {Queyrel}, J., {Tasca}, L., \& {Tresse}, L. 2009, \aap, 504, 789

\bibitem[{{Erb}(2008)}]{Erb08}
{Erb}, D.~K. 2008, \apj, 674, 151

\bibitem[{{Erb} {et~al.}(2006){Erb}, {Steidel}, {Shapley}, {Pettini}, {Reddy},
  \& {Adelberger}}]{Erb+06}
{Erb}, D.~K., {Steidel}, C.~C., {Shapley}, A.~E., {Pettini}, M., {Reddy},
  N.~A., \& {Adelberger}, K.~L. 2006, \apj, 646, 107

\bibitem[{{Finlator} \& {Dav{\'e}}(2008)}]{FinDav08}
{Finlator}, K. \& {Dav{\'e}}, R. 2008, \mnras, 385, 2181

\bibitem[{{F{\"o}rster Schreiber} {et~al.}(2001){F{\"o}rster Schreiber},
  {Genzel}, {Lutz}, {Kunze}, \& {Sternberg}}]{For+01}
{F{\"o}rster Schreiber}, N.~M., {Genzel}, R., {Lutz}, D., {Kunze}, D., \&
  {Sternberg}, A. 2001, \apj, 552, 544

\bibitem[{{F{\"o}rster Schreiber} {et~al.}(2011{\natexlab{a}}){F{\"o}rster
  Schreiber}, {Shapley}, {Erb}, {Genzel}, {Steidel}, {Bouch{\'e}}, {Cresci}, \&
  {Davies}}]{For+11a}
{F{\"o}rster Schreiber}, N.~M., {Shapley}, A.~E., {Erb}, D.~K., {Genzel}, R.,
  {Steidel}, C.~C., {Bouch{\'e}}, N., {Cresci}, G., \& {Davies}, R.
  2011{\natexlab{a}}, \apj, 731, 65

\bibitem[{{F{\"o}rster Schreiber} {et~al.}(2011{\natexlab{b}}){F{\"o}rster
  Schreiber}, {Shapley}, {Genzel}, {Bouch{\'e}}, {Cresci}, {Davies}, {Erb},
  {Genel}, {Lutz}, {Newman}, {Shapiro}, {Steidel}, {Sternberg}, \&
  {Tacconi}}]{For+11b}
{F{\"o}rster Schreiber}, N.~M., {Shapley}, A.~E., {Genzel}, R., {Bouch{\'e}},
  N., {Cresci}, G., {Davies}, R., {Erb}, D.~K., {Genel}, S., {Lutz}, D.,
  {Newman}, S., {Shapiro}, K.~L., {Steidel}, C.~C., {Sternberg}, A., \&
  {Tacconi}, L.~J. 2011{\natexlab{b}}, \apj, 739, 45

\bibitem[{{F{\"o}rster Schreiber} {et~al.}(2009)}]{For+09}
{F{\"o}rster Schreiber}, N.~M. {et~al.} 2009, \apj, 706, 1364

\bibitem[{{Franx} {et~al.}(1997){Franx}, {Illingworth}, {Kelson}, {van Dokkum},
  \& {Tran}}]{Fra+97}
{Franx}, M., {Illingworth}, G.~D., {Kelson}, D.~D., {van Dokkum}, P.~G., \&
  {Tran}, K.-V. 1997, \apjl, 486, L75+

\bibitem[{{Genel} {et~al.}(2010){Genel}, {Naab}, {Genzel}, {F{\"o}rster
  Schreiber}, {Sternberg}, {Oser}, {Johansson}, {Dav{\'e}}, {Oppenheimer}, \&
  {Burkert}}]{Genel+10}
{Genel}, S., {Naab}, T., {Genzel}, R., {F{\"o}rster Schreiber}, N.~M.,
  {Sternberg}, A., {Oser}, L., {Johansson}, P.~H., {Dav{\'e}}, R.,
  {Oppenheimer}, B.~D., \& {Burkert}, A. 2010, ArXiv e-prints

\bibitem[{{Genzel} {et~al.}(2011){Genzel}, {Newman}, {Jones}, {F{\"o}rster
  Schreiber}, {Shapiro}, {Genel}, {Lilly}, {Renzini}, {Tacconi}, {Bouch{\'e}},
  {Burkert}, {Cresci}, {Buschkamp}, {Carollo}, {Ceverino}, {Davies}, {Dekel},
  {Eisenhauer}, {Hicks}, {Kurk}, {Lutz}, {Mancini}, {Naab}, {Peng},
  {Sternberg}, {Vergani}, \& {Zamorani}}]{Gen+11}
{Genzel}, R., {Newman}, S., {Jones}, T., {F{\"o}rster Schreiber}, N.~M.,
  {Shapiro}, K., {Genel}, S., {Lilly}, S.~J., {Renzini}, A., {Tacconi}, L.~J.,
  {Bouch{\'e}}, N., {Burkert}, A., {Cresci}, G., {Buschkamp}, P., {Carollo},
  C.~M., {Ceverino}, D., {Davies}, R., {Dekel}, A., {Eisenhauer}, F., {Hicks},
  E., {Kurk}, J., {Lutz}, D., {Mancini}, C., {Naab}, T., {Peng}, Y.,
  {Sternberg}, A., {Vergani}, D., \& {Zamorani}, G. 2011, \apj, 733, 101

\bibitem[{{Genzel} {et~al.}(2010){Genzel}, {Tacconi}, {Gracia-Carpio},
  {Sternberg}, {Cooper}, {Shapiro}, {Bolatto}, {Bouch{\'e}}, {Bournaud},
  {Burkert}, {Combes}, {Comerford}, {Cox}, {Davis}, {Schreiber},
  {Garcia-Burillo}, {Lutz}, {Naab}, {Neri}, {Omont}, {Shapley}, \&
  {Weiner}}]{Gen+10}
{Genzel}, R., {Tacconi}, L.~J., {Gracia-Carpio}, J., {Sternberg}, A., {Cooper},
  M.~C., {Shapiro}, K., {Bolatto}, A., {Bouch{\'e}}, N., {Bournaud}, F.,
  {Burkert}, A., {Combes}, F., {Comerford}, J., {Cox}, P., {Davis}, M.,
  {Schreiber}, N.~M.~F., {Garcia-Burillo}, S., {Lutz}, D., {Naab}, T., {Neri},
  R., {Omont}, A., {Shapley}, A., \& {Weiner}, B. 2010, \mnras, 407, 2091

\bibitem[{{Genzel} {et~al.}(2006)}]{Gen+06}
{Genzel}, R. {et~al.} 2006, \nat, 442, 786

\bibitem[{{Genzel} {et~al.}(2008)}]{Gen+08}
---. 2008, \apj, 687, 59

\bibitem[{{Gnerucci} {et~al.}(2010){Gnerucci}, {Marconi}, {Capetti}, {Axon}, \&
  {Robinson}}]{Gne+10}
{Gnerucci}, A., {Marconi}, A., {Capetti}, A., {Axon}, D.~J., \& {Robinson}, A.
  2010, \aap, 511, A19+

\bibitem[{{Green} {et~al.}(2010){Green}, {Glazebrook}, {McGregor}, {Abraham},
  {Poole}, {Damjanov}, {McCarthy}, {Colless}, \& {Sharp}}]{Gre+10}
{Green}, A.~W., {Glazebrook}, K., {McGregor}, P.~J., {Abraham}, R.~G., {Poole},
  G.~B., {Damjanov}, I., {McCarthy}, P.~J., {Colless}, M., \& {Sharp}, R.~G.
  2010, \nat, 467, 684

\bibitem[{{Guo} {et~al.}(2011){Guo}, {Giavalisco}, {Ferguson}, {Cassata}, \&
  {Koekemoer}}]{Guo+11}
{Guo}, Y., {Giavalisco}, M., {Ferguson}, H.~C., {Cassata}, P., \& {Koekemoer},
  A.~M. 2011, ArXiv e-prints

\bibitem[{{Heckman} {et~al.}(1990){Heckman}, {Armus}, \& {Miley}}]{HecArmMil90}
{Heckman}, T.~M., {Armus}, L., \& {Miley}, G.~K. 1990, \apjs, 74, 833

\bibitem[{{Heckman} {et~al.}(1993){Heckman}, {Lehnert}, \&
  {Armus}}]{HecLehArm93}
{Heckman}, T.~M., {Lehnert}, M.~D., \& {Armus}, L. 1993, in Astrophysics and
  Space Science Library, Vol. 188, The Environment and Evolution of Galaxies,
  ed. {J.~M.~Shull \& H.~A.~Thronson}, 455--+

\bibitem[{{Hopkins} {et~al.}(2011){Hopkins}, {Quataert}, \& {Murray}}]{Hop+11b}
{Hopkins}, P.~F., {Quataert}, E., \& {Murray}, N. 2011, ArXiv e-prints

\bibitem[{{Hwang} {et~al.}(2011){Hwang}, {Elbaz}, {Dickinson}, {Charmandaris},
  {Daddi}, {Le Borgne}, {Buat}, {Magdis}, {Altieri}, {Aussel}, {Coia},
  {Dannerbauer}, {Dasyra}, {Kartaltepe}, {Leiton}, {Magnelli}, {Popesso}, \&
  {Valtchanov}}]{Hwa+11}
{Hwang}, H.~S., {Elbaz}, D., {Dickinson}, M., {Charmandaris}, V., {Daddi}, E.,
  {Le Borgne}, D., {Buat}, V., {Magdis}, G.~E., {Altieri}, B., {Aussel}, H.,
  {Coia}, D., {Dannerbauer}, H., {Dasyra}, K., {Kartaltepe}, J., {Leiton}, R.,
  {Magnelli}, B., {Popesso}, P., \& {Valtchanov}, I. 2011, ArXiv e-prints

\bibitem[{{Jones} {et~al.}(2010){Jones}, {Ellis}, {Jullo}, \&
  {Richard}}]{Jon+10}
{Jones}, T., {Ellis}, R., {Jullo}, E., \& {Richard}, J. 2010, \apjl, 725, L176

\bibitem[{{Juneau} {et~al.}(2005){Juneau}, {Glazebrook}, {Crampton},
  {McCarthy}, {Savaglio}, {Abraham}, {Carlberg}, {Chen}, {Le Borgne}, {Marzke},
  {Roth}, {J{\o}rgensen}, {Hook}, \& {Murowinski}}]{Jun+05}
{Juneau}, S., {Glazebrook}, K., {Crampton}, D., {McCarthy}, P.~J., {Savaglio},
  S., {Abraham}, R., {Carlberg}, R.~G., {Chen}, H.-W., {Le Borgne}, D.,
  {Marzke}, R.~O., {Roth}, K., {J{\o}rgensen}, I., {Hook}, I., \& {Murowinski},
  R. 2005, \apjl, 619, L135

\bibitem[{{Kassin} {et~al.}(2007){Kassin}, {Weiner}, {Faber}, {Koo}, {Lotz},
  {Diemand}, {Harker}, {Bundy}, {Metevier}, {Phillips}, {Cooper}, {Croton},
  {Konidaris}, {Noeske}, \& {Willmer}}]{Kas+07}
{Kassin}, S.~A., {Weiner}, B.~J., {Faber}, S.~M., {Koo}, D.~C., {Lotz}, J.~M.,
  {Diemand}, J., {Harker}, J.~J., {Bundy}, K., {Metevier}, A.~J., {Phillips},
  A.~C., {Cooper}, M.~C., {Croton}, D.~J., {Konidaris}, N., {Noeske}, K.~G., \&
  {Willmer}, C.~N.~A. 2007, \apjl, 660, L35

\bibitem[{{Katz} {et~al.}(1996){Katz}, {Weinberg}, \& {Hernquist}}]{Kat+96}
{Katz}, N., {Weinberg}, D.~H., \& {Hernquist}, L. 1996, \apjs, 105, 19

\bibitem[{{Kauffmann} {et~al.}(2003){Kauffmann}, {Heckman}, {Tremonti},
  {Brinchmann}, {Charlot}, {White}, {Ridgway}, {Brinkmann}, {Fukugita}, {Hall},
  {Ivezi{\'c}}, {Richards}, \& {Schneider}}]{Kau+03}
{Kauffmann}, G., {Heckman}, T.~M., {Tremonti}, C., {Brinchmann}, J., {Charlot},
  S., {White}, S.~D.~M., {Ridgway}, S.~E., {Brinkmann}, J., {Fukugita}, M.,
  {Hall}, P.~B., {Ivezi{\'c}}, {\v Z}., {Richards}, G.~T., \& {Schneider},
  D.~P. 2003, \mnras, 346, 1055

\bibitem[{{Kennicutt}(1998)}]{Ken98}
{Kennicutt}, Jr., R.~C. 1998, \apj, 498, 541

\bibitem[{{Kennicutt} {et~al.}(2007){Kennicutt}, {Calzetti}, {Walter}, {Helou},
  {Hollenbach}, {Armus}, {Bendo}, {Dale}, {Draine}, {Engelbracht}, {Gordon},
  {Prescott}, {Regan}, {Thornley}, {Bot}, {Brinks}, {de Blok}, {de Mello},
  {Meyer}, {Moustakas}, {Murphy}, {Sheth}, \& {Smith}}]{Ken+07}
{Kennicutt}, Jr., R.~C., {Calzetti}, D., {Walter}, F., {Helou}, G.,
  {Hollenbach}, D.~J., {Armus}, L., {Bendo}, G., {Dale}, D.~A., {Draine},
  B.~T., {Engelbracht}, C.~W., {Gordon}, K.~D., {Prescott}, M.~K.~M., {Regan},
  M.~W., {Thornley}, M.~D., {Bot}, C., {Brinks}, E., {de Blok}, E., {de Mello},
  D., {Meyer}, M., {Moustakas}, J., {Murphy}, E.~J., {Sheth}, K., \& {Smith},
  J.~D.~T. 2007, \apj, 671, 333

\bibitem[{{Kere{\v s}} {et~al.}(2009){Kere{\v s}}, {Katz}, {Fardal},
  {Dav{\'e}}, \& {Weinberg}}]{Ker+09}
{Kere{\v s}}, D., {Katz}, N., {Fardal}, M., {Dav{\'e}}, R., \& {Weinberg},
  D.~H. 2009, \mnras, 395, 160

\bibitem[{{Kewley} \& {Dopita}(2002)}]{KewDop02}
{Kewley}, L.~J. \& {Dopita}, M.~A. 2002, \apjs, 142, 35

\bibitem[{{Kewley} {et~al.}(2001{\natexlab{a}}){Kewley}, {Dopita},
  {Sutherland}, {Heisler}, \& {Trevena}}]{Kew+01b}
{Kewley}, L.~J., {Dopita}, M.~A., {Sutherland}, R.~S., {Heisler}, C.~A., \&
  {Trevena}, J. 2001{\natexlab{a}}, \apj, 556, 121

\bibitem[{{Kewley} {et~al.}(2001{\natexlab{b}}){Kewley}, {Heisler}, {Dopita},
  \& {Lumsden}}]{Kew+01a}
{Kewley}, L.~J., {Heisler}, C.~A., {Dopita}, M.~A., \& {Lumsden}, S.
  2001{\natexlab{b}}, \apjs, 132, 37

\bibitem[{{Komatsu} {et~al.}(2011){Komatsu}, {Smith}, {Dunkley}, {Bennett},
  {Gold}, {Hinshaw}, {Jarosik}, {Larson}, {Nolta}, {Page}, {Spergel},
  {Halpern}, {Hill}, {Kogut}, {Limon}, {Meyer}, {Odegard}, {Tucker}, {Weiland},
  {Wollack}, \& {Wright}}]{Kom+11}
{Komatsu}, E., {Smith}, K.~M., {Dunkley}, J., {Bennett}, C.~L., {Gold}, B.,
  {Hinshaw}, G., {Jarosik}, N., {Larson}, D., {Nolta}, M.~R., {Page}, L.,
  {Spergel}, D.~N., {Halpern}, M., {Hill}, R.~S., {Kogut}, A., {Limon}, M.,
  {Meyer}, S.~S., {Odegard}, N., {Tucker}, G.~S., {Weiland}, J.~L., {Wollack},
  E., \& {Wright}, E.~L. 2011, \apjs, 192, 18

\bibitem[{{Law} {et~al.}(2009){Law}, {Steidel}, {Erb}, {Larkin}, {Pettini},
  {Shapley}, \& {Wright}}]{Law+09}
{Law}, D.~R., {Steidel}, C.~C., {Erb}, D.~K., {Larkin}, J.~E., {Pettini}, M.,
  {Shapley}, A.~E., \& {Wright}, S.~A. 2009, \apj, 697, 2057

\bibitem[{{Law} {et~al.}(2011){Law}, {Steidel}, {Shapley}, {Nagy}, {Reddy}, \&
  {Erb}}]{Law+11}
{Law}, D.~R., {Steidel}, C.~C., {Shapley}, A.~E., {Nagy}, S.~R., {Reddy},
  N.~A., \& {Erb}, D.~K. 2011, ArXiv e-prints

\bibitem[{{Le Borgne} {et~al.}(2009){Le Borgne}, {Elbaz}, {Ocvirk}, \&
  {Pichon}}]{Leb+09}
{Le Borgne}, D., {Elbaz}, D., {Ocvirk}, P., \& {Pichon}, C. 2009, \aap, 504,
  727

\bibitem[{{Le Floc'h} {et~al.}(2005){Le Floc'h}, {Papovich}, {Dole}, {Bell},
  {Lagache}, {Rieke}, {Egami}, {P{\'e}rez-Gonz{\'a}lez}, {Alonso-Herrero},
  {Rieke}, {Blaylock}, {Engelbracht}, {Gordon}, {Hines}, {Misselt}, {Morrison},
  \& {Mould}}]{Lef+05}
{Le Floc'h}, E., {Papovich}, C., {Dole}, H., {Bell}, E.~F., {Lagache}, G.,
  {Rieke}, G.~H., {Egami}, E., {P{\'e}rez-Gonz{\'a}lez}, P.~G.,
  {Alonso-Herrero}, A., {Rieke}, M.~J., {Blaylock}, M., {Engelbracht}, C.~W.,
  {Gordon}, K.~D., {Hines}, D.~C., {Misselt}, K.~A., {Morrison}, J.~E., \&
  {Mould}, J. 2005, \apj, 632, 169

\bibitem[{{Lehnert} \& {Heckman}(1996)}]{LehHec96}
{Lehnert}, M.~D. \& {Heckman}, T.~M. 1996, \apj, 462, 651

\bibitem[{{Lehnert} {et~al.}(1999){Lehnert}, {Heckman}, \& {Weaver}}]{Leh+99}
{Lehnert}, M.~D., {Heckman}, T.~M., \& {Weaver}, K.~A. 1999, \apj, 523, 575

\bibitem[{{Leitherer} {et~al.}(1999){Leitherer}, {Schaerer}, {Goldader},
  {Gonz{\'a}lez Delgado}, {Robert}, {Kune}, {de Mello}, {Devost}, \&
  {Heckman}}]{Lei+99}
{Leitherer}, C., {Schaerer}, D., {Goldader}, J.~D., {Gonz{\'a}lez Delgado},
  R.~M., {Robert}, C., {Kune}, D.~F., {de Mello}, D.~F., {Devost}, D., \&
  {Heckman}, T.~M. 1999, \apjs, 123, 3

\bibitem[{{Leroy} {et~al.}(2012){Leroy}, {Bigiel}, {de Blok}, {Boissier},
  {Bolatto}, {Brinks}, {Madore}, {Munoz-Mateos}, {Murphy}, {Sandstrom},
  {Schruba}, \& {Walter}}]{Ler+12}
{Leroy}, A.~K., {Bigiel}, F., {de Blok}, W.~J.~G., {Boissier}, S., {Bolatto},
  A., {Brinks}, E., {Madore}, B., {Munoz-Mateos}, J.-C., {Murphy}, E.,
  {Sandstrom}, K., {Schruba}, A., \& {Walter}, F. 2012, ArXiv e-prints

\bibitem[{{Levesque} {et~al.}(2010){Levesque}, {Kewley}, \& {Larson}}]{Lev+10}
{Levesque}, E.~M., {Kewley}, L.~J., \& {Larson}, K.~L. 2010, \aj, 139, 712

\bibitem[{{Lilly} {et~al.}(1996){Lilly}, {Le Fevre}, {Hammer}, \&
  {Crampton}}]{Lil+96}
{Lilly}, S.~J., {Le Fevre}, O., {Hammer}, F., \& {Crampton}, D. 1996, \apjl,
  460, L1+

\bibitem[{{Lilly} {et~al.}(2007)}]{Lil+07}
{Lilly}, S.~J. {et~al.} 2007, \apjs, 172, 70

\bibitem[{{Liu} {et~al.}(2008){Liu}, {Shapley}, {Coil}, {Brinchmann}, \&
  {Ma}}]{Liu+08}
{Liu}, X., {Shapley}, A.~E., {Coil}, A.~L., {Brinchmann}, J., \& {Ma}, C.-P.
  2008, \apj, 678, 758

\bibitem[{{Madau} {et~al.}(1996){Madau}, {Ferguson}, {Dickinson}, {Giavalisco},
  {Steidel}, \& {Fruchter}}]{Mad+96}
{Madau}, P., {Ferguson}, H.~C., {Dickinson}, M.~E., {Giavalisco}, M.,
  {Steidel}, C.~C., \& {Fruchter}, A. 1996, \mnras, 283, 1388

\bibitem[{{Mancini} {et~al.}(2011){Mancini}, {Foerster Schreiber}, {Renzini},
  {Cresci}, {Hicks}, {Peng}, {Vergani}, {Lilly}, {Carollo}, {Pozzetti},
  {Zamorani}, {Daddi}, {Genzel}, {Maraston}, {McCracken}, {Tacconi}, {Bouche},
  {Davies}, {Oesch}, {Shapiro}, {Mainieri}, {Lutz}, {Mignoli}, \&
  {Sternberg}}]{Man+11}
{Mancini}, C., {Foerster Schreiber}, N., {Renzini}, A., {Cresci}, G., {Hicks},
  E., {Peng}, Y., {Vergani}, D., {Lilly}, S., {Carollo}, C.~M., {Pozzetti}, L.,
  {Zamorani}, G., {Daddi}, E., {Genzel}, R., {Maraston}, C., {McCracken},
  H.~J., {Tacconi}, L.~J., {Bouche}, N., {Davies}, R.~I., {Oesch}, P.,
  {Shapiro}, K., {Mainieri}, V., {Lutz}, D., {Mignoli}, M., \& {Sternberg}, A.
  2011, ArXiv e-prints

\bibitem[{{Martin}(1998)}]{Mar98}
{Martin}, C.~L. 1998, \apj, 506, 222

\bibitem[{{Martin}(1999)}]{Mar99}
---. 1999, \apj, 513, 156

\bibitem[{{Martin}(2005)}]{Mar05}
---. 2005, \apj, 621, 227

\bibitem[{{Matzner}(2002)}]{Mat02}
{Matzner}, C.~D. 2002, \apj, 566, 302

\bibitem[{{McCracken} {et~al.}(2010){McCracken}, {Capak}, {Salvato}, {Aussel},
  {Thompson}, {Daddi}, {Sanders}, {Kneib}, {Willott}, {Mancini}, {Renzini},
  {Cook}, {Le F{\`e}vre}, {Ilbert}, {Kartaltepe}, {Koekemoer}, {Mellier},
  {Murayama}, {Scoville}, {Shioya}, \& {Tanaguchi}}]{McCra+10}
{McCracken}, H.~J., {Capak}, P., {Salvato}, M., {Aussel}, H., {Thompson}, D.,
  {Daddi}, E., {Sanders}, D.~B., {Kneib}, J.-P., {Willott}, C.~J., {Mancini},
  C., {Renzini}, A., {Cook}, R., {Le F{\`e}vre}, O., {Ilbert}, O.,
  {Kartaltepe}, J., {Koekemoer}, A.~M., {Mellier}, Y., {Murayama}, T.,
  {Scoville}, N.~Z., {Shioya}, Y., \& {Tanaguchi}, Y. 2010, \apj, 708, 202

\bibitem[{{McKee} \& {Ostriker}(1977)}]{McKeeOst77}
{McKee}, C.~F. \& {Ostriker}, J.~P. 1977, \apj, 218, 148

\bibitem[{{Murray} {et~al.}(2005){Murray}, {Quataert}, \& {Thompson}}]{Mur+05}
{Murray}, N., {Quataert}, E., \& {Thompson}, T.~A. 2005, \apj, 618, 569

\bibitem[{{Murray} {et~al.}(2010){Murray}, {Quataert}, \& {Thompson}}]{Mur+10}
---. 2010, \apj, 709, 191

\bibitem[{{Osterbrock}(1989)}]{Ost89}
{Osterbrock}, D.~E. 1989, {Astrophysics of gaseous nebulae and active galactic
  nuclei} ({University Science Books})

\bibitem[{{Ostriker} \& {Shetty}(2011)}]{OstShe11}
{Ostriker}, E.~C. \& {Shetty}, R. 2011, \apj, 731, 41

\bibitem[{{Pettini} {et~al.}(2001){Pettini}, {Shapley}, {Steidel}, {Cuby},
  {Dickinson}, {Moorwood}, {Adelberger}, \& {Giavalisco}}]{Pet+01}
{Pettini}, M., {Shapley}, A.~E., {Steidel}, C.~C., {Cuby}, J.-G., {Dickinson},
  M., {Moorwood}, A.~F.~M., {Adelberger}, K.~L., \& {Giavalisco}, M. 2001,
  \apj, 554, 981

\bibitem[{{Pettini} {et~al.}(2000){Pettini}, {Steidel}, {Adelberger},
  {Dickinson}, \& {Giavalisco}}]{Pet+00}
{Pettini}, M., {Steidel}, C.~C., {Adelberger}, K.~L., {Dickinson}, M., \&
  {Giavalisco}, M. 2000, \apj, 528, 96

\bibitem[{{Rich} {et~al.}(2011){Rich}, {Kewley}, \& {Dopita}}]{Ric+11}
{Rich}, J.~A., {Kewley}, L.~J., \& {Dopita}, M.~A. 2011, \apj, 734, 87

\bibitem[{{Rubin} {et~al.}(2010){Rubin}, {Weiner}, {Koo}, {Martin},
  {Prochaska}, {Coil}, \& {Newman}}]{Rub+10}
{Rubin}, K.~H.~R., {Weiner}, B.~J., {Koo}, D.~C., {Martin}, C.~L., {Prochaska},
  J.~X., {Coil}, A.~L., \& {Newman}, J.~A. 2010, \apj, 719, 1503

\bibitem[{{Rupke} {et~al.}(2005){Rupke}, {Veilleux}, \&
  {Sanders}}]{RupVeiSan05}
{Rupke}, D.~S., {Veilleux}, S., \& {Sanders}, D.~B. 2005, \apjs, 160, 115

\bibitem[{{Rupke} \& {Veilleux}(2011)}]{RupVei11}
{Rupke}, D.~S.~N. \& {Veilleux}, S. 2011, \apjl, 729, L27+

\bibitem[{{Sargent} {et~al.}(2012){Sargent}, {B{\'e}thermin}, {Daddi}, \&
  {Elbaz}}]{Sar+12}
{Sargent}, M.~T., {B{\'e}thermin}, M., {Daddi}, E., \& {Elbaz}, D. 2012, \apjl,
  747, L31

\bibitem[{{Sato} {et~al.}(2009){Sato}, {Martin}, {Noeske}, {Koo}, \&
  {Lotz}}]{Sat+09}
{Sato}, T., {Martin}, C.~L., {Noeske}, K.~G., {Koo}, D.~C., \& {Lotz}, J.~M.
  2009, \apj, 696, 214

\bibitem[{{Schreiber} {et~al.}(2004){Schreiber}, {Thatte}, {Eisenhauer},
  {Tecza}, {Abuter}, \& {Horrobin}}]{Sch+04}
{Schreiber}, J., {Thatte}, N., {Eisenhauer}, F., {Tecza}, M., {Abuter}, R., \&
  {Horrobin}, M. 2004, in Astronomical Society of the Pacific Conference
  Series, Vol. 314, Astronomical Data Analysis Software and Systems (ADASS)
  XIII, ed. {F.~Ochsenbein, M.~G.~Allen, \& D.~Egret}, 380

\bibitem[{{Shapiro} {et~al.}(2008)}]{Sha+08}
{Shapiro}, K.~L. {et~al.} 2008, \apj, 682, 231

\bibitem[{{Shapiro} {et~al.}(2009)}]{Sha+09}
---. 2009, \apj, 701, 955

\bibitem[{{Shapley} {et~al.}(2005){Shapley}, {Steidel}, {Erb}, {Reddy},
  {Adelberger}, {Pettini}, {Barmby}, \& {Huang}}]{Shapley+05}
{Shapley}, A.~E., {Steidel}, C.~C., {Erb}, D.~K., {Reddy}, N.~A., {Adelberger},
  K.~L., {Pettini}, M., {Barmby}, P., \& {Huang}, J. 2005, \apj, 626, 698

\bibitem[{{Shapley} {et~al.}(2003){Shapley}, {Steidel}, {Pettini}, \&
  {Adelberger}}]{Shapley+03}
{Shapley}, A.~E., {Steidel}, C.~C., {Pettini}, M., \& {Adelberger}, K.~L. 2003,
  \apj, 588, 65

\bibitem[{{Somerville} \& {Primack}(1999)}]{SomPri99}
{Somerville}, R.~S. \& {Primack}, J.~R. 1999, \mnras, 310, 1087

\bibitem[{{Springel} \& {Hernquist}(2003)}]{SprHer03}
{Springel}, V. \& {Hernquist}, L. 2003, \mnras, 339, 312

\bibitem[{{Stahler} \& {Palla}(2005)}]{StaPal05}
{Stahler}, S.~W. \& {Palla}, F. 2005, {The Formation of Stars} (Wiley-VCH)

\bibitem[{{Steidel} {et~al.}(1999){Steidel}, {Adelberger}, {Giavalisco},
  {Dickinson}, \& {Pettini}}]{Ste+99}
{Steidel}, C.~C., {Adelberger}, K.~L., {Giavalisco}, M., {Dickinson}, M., \&
  {Pettini}, M. 1999, \apj, 519, 1

\bibitem[{{Steidel} {et~al.}(2010){Steidel}, {Erb}, {Shapley}, {Pettini},
  {Reddy}, {Bogosavljevi{\'c}}, {Rudie}, \& {Rakic}}]{Ste+10}
{Steidel}, C.~C., {Erb}, D.~K., {Shapley}, A.~E., {Pettini}, M., {Reddy}, N.,
  {Bogosavljevi{\'c}}, M., {Rudie}, G.~C., \& {Rakic}, O. 2010, \apj, 717, 289

\bibitem[{{Steidel} {et~al.}(1996){Steidel}, {Giavalisco}, {Pettini},
  {Dickinson}, \& {Adelberger}}]{Ste+96}
{Steidel}, C.~C., {Giavalisco}, M., {Pettini}, M., {Dickinson}, M., \&
  {Adelberger}, K.~L. 1996, \apjl, 462, L17+

\bibitem[{{Strickland} \& {Heckman}(2009)}]{StrHec09}
{Strickland}, D.~K. \& {Heckman}, T.~M. 2009, \apj, 697, 2030

\bibitem[{{Strickland} {et~al.}(2004){Strickland}, {Heckman}, {Colbert},
  {Hoopes}, \& {Weaver}}]{Str+04}
{Strickland}, D.~K., {Heckman}, T.~M., {Colbert}, E.~J.~M., {Hoopes}, C.~G., \&
  {Weaver}, K.~A. 2004, \apjs, 151, 193

\bibitem[{{Sturm} {et~al.}(2011){Sturm}, {Gonz{\'a}lez-Alfonso}, {Veilleux},
  {Fischer}, {Graci{\'a}-Carpio}, {Hailey-Dunsheath}, {Contursi}, {Poglitsch},
  {Sternberg}, {Davies}, {Genzel}, {Lutz}, {Tacconi}, {Verma}, {Maiolino}, \&
  {de Jong}}]{Stu+11}
{Sturm}, E., {Gonz{\'a}lez-Alfonso}, E., {Veilleux}, S., {Fischer}, J.,
  {Graci{\'a}-Carpio}, J., {Hailey-Dunsheath}, S., {Contursi}, A., {Poglitsch},
  A., {Sternberg}, A., {Davies}, R., {Genzel}, R., {Lutz}, D., {Tacconi}, L.,
  {Verma}, A., {Maiolino}, R., \& {de Jong}, J.~A. 2011, \apjl, 733, L16+

\bibitem[{{Sutherland} \& {Dopita}(1993)}]{SutDop93}
{Sutherland}, R.~S. \& {Dopita}, M.~A. 1993, \apjs, 88, 253

\bibitem[{{Swinbank} {et~al.}(2011){Swinbank}, {Papadopoulos}, {Cox}, {Krips},
  {Ivison}, {Smail}, {Thomson}, {Neri}, {Richard}, \& {Ebeling}}]{Swi+11}
{Swinbank}, A.~M., {Papadopoulos}, P.~P., {Cox}, P., {Krips}, M., {Ivison},
  R.~J., {Smail}, I., {Thomson}, A.~P., {Neri}, R., {Richard}, J., \&
  {Ebeling}, H. 2011, \apj, 742, 11

\bibitem[{{Tacconi} {et~al.}(2008)}]{Tac+08}
{Tacconi}, L.~J. {et~al.} 2008, \apj, 680, 246

\bibitem[{{Tacconi} {et~al.}(2010)}]{Tac+10}
---. 2010, \nat, 463, 781

\bibitem[{{Veilleux} {et~al.}(2005){Veilleux}, {Cecil}, \&
  {Bland-Hawthorn}}]{Vei+05}
{Veilleux}, S., {Cecil}, G., \& {Bland-Hawthorn}, J. 2005, \araa, 43, 769

\bibitem[{{Veilleux} \& {Osterbrock}(1987)}]{VeiOst87}
{Veilleux}, S. \& {Osterbrock}, D.~E. 1987, \apjs, 63, 295

\bibitem[{{Weiner} {et~al.}(2009)}]{Wei+09}
{Weiner}, B.~J. {et~al.} 2009, \apj, 692, 187

\bibitem[{{Wuyts} {et~al.}(2011{\natexlab{a}}){Wuyts}, {F{\"o}rster Schreiber},
  {Lutz}, {Nordon}, {Berta}, {Altieri}, {Andreani}, {Aussel}, {Bongiovanni},
  {Cepa}, {Cimatti}, {Daddi}, {Elbaz}, {Genzel}, {Koekemoer}, {Magnelli},
  {Maiolino}, {McGrath}, {P{\'e}rez Garc{\'{\i}}a}, {Poglitsch}, {Popesso},
  {Pozzi}, {Sanchez-Portal}, {Sturm}, {Tacconi}, \& {Valtchanov}}]{Wuy+11a}
{Wuyts}, S., {F{\"o}rster Schreiber}, N.~M., {Lutz}, D., {Nordon}, R., {Berta},
  S., {Altieri}, B., {Andreani}, P., {Aussel}, H., {Bongiovanni}, A., {Cepa},
  J., {Cimatti}, A., {Daddi}, E., {Elbaz}, D., {Genzel}, R., {Koekemoer},
  A.~M., {Magnelli}, B., {Maiolino}, R., {McGrath}, E.~J., {P{\'e}rez
  Garc{\'{\i}}a}, A., {Poglitsch}, A., {Popesso}, P., {Pozzi}, F.,
  {Sanchez-Portal}, M., {Sturm}, E., {Tacconi}, L., \& {Valtchanov}, I.
  2011{\natexlab{a}}, \apj, 738, 106

\bibitem[{{Wuyts} {et~al.}(2011{\natexlab{b}}){Wuyts}, {Forster Schreiber},
  {van der Wel}, {Magnelli}, {Guo}, {Genzel}, {Lutz}, {Aussel}, {Berta},
  {Cava}, {Gracia-Carpio}, {Kocevski}, {Koekemoer}, {Lee}, {Le Floc'h},
  {McGrath}, {Nordon}, {Popesso}, {Pozzi}, {Riguccini}, {Rodighiero},
  {Saintonge}, \& {Tacconi}}]{Wuy+11b}
{Wuyts}, S., {Forster Schreiber}, N.~M., {van der Wel}, A., {Magnelli}, B.,
  {Guo}, Y., {Genzel}, R., {Lutz}, D., {Aussel}, H., {Berta}, S., {Cava}, A.,
  {Gracia-Carpio}, J., {Kocevski}, D.~D., {Koekemoer}, A.~M., {Lee}, K.-S., {Le
  Floc'h}, E., {McGrath}, E.~J., {Nordon}, R., {Popesso}, P., {Pozzi}, F.,
  {Riguccini}, L., {Rodighiero}, G., {Saintonge}, A., \& {Tacconi}, L.
  2011{\natexlab{b}}, ArXiv e-prints

\end{thebibliography}

\begin{table}[h]
\caption{Emission Line Data from the Clumps}
\begin{center}
\begin{tabular}{l l l l l l l l}
\hline
Region & \% broad & $\sigma_{narrow}$ & $\sigma_{broad}$ & \niiha & \siiha & \% shock$^{a}$ & $\chi^{2}$ fit \\
 \hline
 \hline
 Clump A & 50 $\pm$ 4.8 & 82.7 $\pm$ 4.2 & 199 $\pm$11.9 & 0.11 $\pm$ 0.014 & 0.14 $\pm$ 0.029 & 5 $\pm$ 5 & 0.78 \\
 \hline
 Wind A & 44 $\pm$ 7.2 & 84.7 $\pm$ 4.9 & 197 $\pm$ 20.5 & 0.15 $\pm$ 0.036 & 0.21 $\pm$ 0.061 & 10 $^{+10}_{-5}$ & 0.64\\
 \hline
 Clump B & 46 $\pm$ 4.6 & 81.9 $\pm$ 4.6 & 244 $\pm$ 68.2 & 0.20 $\pm$ 0.031 & 0.20 $\pm$ 0.055 & 15 $^{+25}_{-15}$ & 0.59 \\
 \hline
 Wind B & 71 $\pm$ 2.9 & 97.7 $\pm$ 6.0 & 291 $\pm$ 12.7 & 0.32 $\pm$ 0.032 & 0.27 $\pm$ 0.068 & 30 $^{+30}_{-20}$ & 0.58 \\ 
 \hline
 \end{tabular}
  \end{center}

$^{a}$The shock contribution is calculated by comparing the data to Figure 10 of \cite{Ric+11}.
 \end{table}
 
 \begin{table}[h]
\caption{Observed Properties of the Clumps}
\begin{center}
\begin{tabular}{l l l}
\hline
& Clump A & Clump B \\ 
 \hline
 \hline
 E(B-V)$^{a}$ (H$\alpha$/H$\beta$) & 0.493 $\pm$ 0.165 & 0.493$\pm$ 0.165 \\ 
 \hline
 E(B-V) (SED) & 0.33 $\pm$ 0.1 & 0.33 $\pm$ 0.1 \\
 \hline
 \oii\lam3729/3726 & 1.38 $\pm$ 0.41 & 1.51 $\pm$ 1.14 \\
 \hline
 n$_{e}$ (\oii) (cm$^{-3}$)$^{b}$ & 80, [1--100]$^{c}$ & 5, [1--1000]$^{c}$ \\
 \hline
 \sii\lam6716/6731$^{d}$ & 0.7 $\pm$ 0.1 & 1.2 $\pm$ 0.3 \\
 \hline
  n$_{e}$ (\sii) (cm$^{-3}$)$^{b,d}$ & 1800, [1100--3000] $^{c}$ & 290, [30-750]$^{c}$ \\
  \hline
  \oiiihb$^{a}$ & 6.72 $\pm$ 1.31 & 4.62 $\pm$ 10.30 \\
  \hline
  \niiha$^{e}$ & 0.11 $\pm$ 0.018 & 0.21 $\pm$ 0.029 \\
  \hline
  \siiha$^{e}$ & 0.18 $\pm$ 0.049 & 0.18 $\pm$ 0.067 \\
  \hline
  \oii/\oiii$^{e}$ & 0.99 $\pm$ 0.14 & 1.06 $\pm$ 0.35\\
  \hline
  U (from O$_{32}$) (cm$^{-1}$)$^{a,g} $ & 5.50x10$^{7}$, [4.75x10$^{7}$--1.1x10$^{8}$]$^{c}$ & 8.75x10$^{7}$, [3.75x10$^{7}$--1.2x10$^{8}$]$^{c}$ \\
  \hline
  U (from O$_{32}$) (cm$^{-1}$)$^{f,g}$ & ... & 1.03x10$^{8}$, [4.50x10$^{7}$--1.25x10$^{8}$]$^{c}$ \\
  \hline
  SFR (\msunyr)$^{f}$ & 40 $\pm$ 12 & 11 $\pm$ 3.3 \\
  \hline
  r$_{out,HWHM}$ (kpc)$^{f}$ & 0.8 $\pm$ 0.3 & 1.2 $\pm$ 0.5 \\
  \hline
  U (from SFR) (cms$^{-1}$) & 7.4x10$^{7}$ $\pm$ 6.1x10$^{7}$ & 8.3x10$^{7}$ $\pm$ 1.0x10$^{8}$ \\
  \hline
  12 + log10(O/H)$^{h}$ & 8.42 $\pm$ 0.12 & 8.62 $\pm$ 0.096\\
  \hline
  \end{tabular}
   \end{center}

  $^{a}$ For clump B, these numbers are estimated assuming \hahb is the same for clumps A and B.
  
  $^{b}$ n$_{e}$ is calculated from \cite{Ost89} assuming collisional de-excitation with T$_{e}$ = 10$^{4}$ K.
  
  $^{c}$ 1$\sigma$ range
  
$^{d}$ from \cite{Gen+11}

$^{e}$ For regions including both the clump and the wind which are spatially consistent among H, J, and K-band cubes.

$^{f}$ For clump B, assuming \hahb is the same as for the diffuse gas.

$^{g}$ For this calculation, metallicities are assumed to be those as calculated using the combined method of \cite{KewDop02}.

$^{h}$ Using the \niiha conversion of \cite{deNic+02}.
 \end{table}
 
 \begin{table}[h]
 \caption{Calculated Properties of the Clumps and their Outflows}
 \begin{center}
 \begin{tabular}{l l l}
 \hline
  & Clump A & Clump B \\
  \hline
  \hline
  SFR (\msunyr)$^{a}$ & 40 $\pm$ 12 & 11 $\pm$ 3.3 \\
  \hline
  \mdotout (\msunyr)$^{a}$ & 117 $^{+1110}_{-95}$ & 78 $^{+590}_{-78}$ \\
  \hline
  \vout (\kms)$^{a}$ & 440 $\pm$ 150 & 810 $\pm$ 270\\
  \hline
  \vout/v$_{escape,clump}$ & 0.75 $\pm$ 0.39 & 2.4 $\pm$ 1.3 \\
  \hline
  \vout/v$_{escape,galaxy}$ & 0.54 $\pm$ 0.26 & 0.99 $\pm$ 0.48 \\
  \hline
  t$_{dyn}$ (Myr) & 1.8 $\pm$ 0.9 & 1.5 $\pm$ 0.8 \\
  \hline
  \Mw (\msun) & 1.79x10$^{8}$ $^{+1.96\times10^{9}}_{-1.43\times10^{8}}$& 1.09x10$^{8}$ $^{+8.36\times10^{8}}_{-9.45\times10^{7}}$\\
  \hline
  \Ew (ergs) & 3.46x10$^{56}$ $^{+3.81\times10^{57}}_{-3.21\times10^{56}}$ & 7.16x10$^{56}$ $^{+5.50\times10^{57}}_{-7.06\times10^{56}}$\\
  \hline
  M$_{gas,mol}$ (\msun) & 1.6x10$^{10}$ $^{+1.1\times10^{10}}_{-1.3\times10^{10}}$ & 7.8x10$^{9}$ $^{+5.5\times10^{9}}_{-6.6\times10^{9}}$ \\
  \hline
  \siggas (\msunpc)$^{a}$ & 8000 $^{+5600}_{-6800}$ & 1200 $^{+840}_{-1000}$ \\
  \hline
  $\Sigma_{SFR}$ (\msunpc yr$^{-1}$)$^{a}$ & 13.5 $\pm$ 4.0 & 1.6 $\pm$ 0.5 \\
  \hline
  \Pkin (dyne cm$^{-2}$) & 2.9x10$^{-7}$ $\pm$ 1.6x10$^{-7}$ & 5.3x10$^{-8}$ $\pm$ 5.5x10$^{-8}$ \\
  \hline
  \Pw (dyne cm$^{-2}$) & 3.0x10$^{-7}$  $^{+2.9\times10^{-7}}_{-3.6\times10^{-7}}$ & 6.7x10$^{-9}$  $^{+6.6\times10^{-9}}_{-8.0\times10^{-9}}$\\
  \hline
  \Pturb (dyne cm$^{-2}$) & 6.8x10$^{-9}$ ($\pm$ 2.1x10$^{-9}$) x f$_{p}$ & 8.4x10$^{-10}$ ($\pm$ 2.5x10$^{-10}$) x f$_{p}$\\
  \hline
  \Prad (dyne cm$^{-2}$) & 7.1x10$^{-9}$ $^{+5.4\times10^{-9}}_{-6.4\times10^{-9}}$ & 1.3x10$^{-10}$ $^{+1.0\times10^{-10}}_{-1.2\times10^{-10}}$ \\
  \hline
  P$_{ram}$ (dyne cm$^{-2}$) & 4.4x10$^{-7}$ ($^{+1.8\times10^{-6}}_{-4.0\times10^{-7}}$) x ($\rho_{w}$/100) & 1.5x10$^{-6}$ ($^{+6.0\times10^{-6}}_{-1.3\times10^{-6}}$) x ($\rho_{w}$/100) \\
  \hline
  $\rho_{wind}$ (cm$^{-3}$)$^{b}$ & 67.3 $\pm$ 73.7 & 0.4 $\pm$ 0.5 \\
  \hline
  L$_{bol}$ (ergss$^{-1}$) & 1.6x10$^{45}$ $\pm$ 4.8x10$^{44}$ & 4.4x10$^{44}$ $\pm$1.3x10$^{44}$ \\
  \hline
  \mdotout x \vout$^{2}$/ L$_{bol}$ & 0.0045, [7.64x10$^{-5}$--0.048]$^{c}$ & 0.046, [0.00021--0.33]$^{c}$  \\
  \hline
  \pdotout/P$_{rad}$ & 6.2, [0.5--65.1]$^{c}$ & 33.7, [2.1--244.5]$^{c}$ \\
  \hline
  $\eta$ (\mdotout/SFR)$^{a}$ & 2.9 $^{+27.6}_{-2.5}$ & 8.6 $^{+53.7}_{-7.5}$ \\
  \hline
  E$_{shock}$ (ergs) & 3.46 x10$^{55}$ $^{+3.81\times10^{56}}_{-3.65\times10^{55}}$ & 2.15x10$^{56}$ $^{+1.66\times10^{57}}_{-3.86\times10^{56}}$ \\
  \hline
  E$_{turb}$ (ergs) & 3.28x10$^{57}$ $^{+2.31\times10^{57}}_{-2.77\times10^{57}}$ & 1.89x10$^{57}$ $^{+1.33\times10^{57}}_{-1.59\times10^{57}}$ \\
  \hline
  f$_{v}$ (n$_{avg}$/n$_{local}$) & 2.2x10$^{-2}$--3.7x10$^{-4}$ & 2.3x10$^{-3}$--2.2x10$^{-4}$  \\
  \hline
  \end{tabular}
     \end{center}
$^{a}$ from \cite{Gen+11}

$^{b}$ assuming P$_{ram} \sim$ \Pw

$^{c}$ 1$\sigma$ range

 \end{table}
 
 \begin{table}[h]
 \caption{Comparison of Clump and Galaxy Properties}
 \begin{center}
 \begin{tabular}{l l l l l}
 \hline
 & Clumps & ZC406690 & Milky Way & ULIRGs \\
 \hline
 \hline
 SFR (\msunyr) & 10--40 & 300 & 3 & 100--1000 \\
 \hline
 \siggas (\msunpc) & 1200--8000 & 1100 & 200 & 3000--30,000 \\
 \hline
 f$_{gas}$ & 0.4--0.7 & 0.3 & 0.1 & 0.1 \\
 \hline
 log10(P/k)$^{a}$ & 8.6--8.9 & 9.0 & 5.9 & 9.3 \\
 \hline
 log10(n$_{H_{2}}$) & 4.0--5.8 & ... & 2 & 3--4 \\
 \hline
 $\sigma$ (\kms) & 85 & 63 & 5 & 50--100 \\
 \hline
 log10(U) (cms$^{-1}$) & 7.8--8 & ... & 7.5 & 7.7 \\
 \hline
$\eta$$^{b}$ & 3--8 & ... & ... & 0.1--2.5 \\
\hline
P/(L$_{bol}$/c)$^{b}$ & 6--33 & ... & ... & 0.1--0.5 \\
\hline
$\frac{dE}{dt}$/L$_{IR}$$^{b}$ & 0.04-0.005 & ... & ... & 0.0001-0.007 \\
\hline
f$_{v}$ (n$_{avg}$/n$_{local}$) & 2x10$^{-2}$--4x10$^{-4}$ & ... & 10$^{-2}$ & 3.6x10$^{-3}$ \\
\hline
\end{tabular} 
\end{center}

$^{a}$ from \Pkin 

$^{b}$These values are derived from \ha for ZC406690 and from Na ID for ULIRGs \citep{RupVeiSan05,RupVei11}. However, this might not be a fair comparison since these measurements sample very different components of the wind. In addition, the largest values for ULIRGs (for $\eta$ and $\frac{dE}{dt}$/L$_{IR}$), which come from Mrk231, are lower limits \citep{RupVei11}. \\
\end{table}

\pagebreak

\appendix
\section{Appendix A: Systematic errors for M$_{gas}$}
In this section, we derive the systematic errors in our M$_{gas}$ calculation from a combination of (1) scatter in the intrinsic relation, (2) the varying slope and zero point of the relation as quoted by different authors, (3) the extrapolation of the relation from low-z to high-z, and (4) the application of the relation on $\sim$kpc scales. 

First, we consider the scatter of \siggas$~$ within an individual Kennicutt-Schmidt (KS) relation. \cite{Big+11} compare \siggas$~$ and $\Sigma_{SFR}$ for thousands of positions in 30 nearby disk galaxies, and find that the 1$\sigma$ scatter is roughly 0.3 dex or 70\% (see their Figure 1). \cite{Gen+10} find a similar systematic uncertainty for \siggas$~$ of 0.3 dex in the KS relation for both local and high-z galaxies. 

Second, we consider the systematic uncertainty resulting from extrapolating the low-z KS relation to a high-z galaxy. As seen in \cite{Gen+10} and \cite{Dad+10}, high and low-z galaxies on the star-forming Òmain sequenceÓ (that is, neglecting local ULIRGs and high-z SMGs) fit on the same KS relation with an offset in zero point of less than a few x 0.1 dex, in the sense that high-z galaxies could have slightly lower gas masses than the local relation would predict. ZC406690 falls on the upper SFR and M$_{*}$ end of the SF main-sequence.

Third, we consider the scatter between different versions of the KS relation by looking at the most recent observational results \citep{Gen+10,Dad+10,Ken+07}. We do not consider the slope and zero-point of the \cite{Big+11} relation, as they note that theirs is not a ÒrigorousÓ fit. From the three other relations, the maximum difference in \siggas$~$ for our clumps is around 50\%. Both the \cite{Ken+07} and the \cite{Dad+10} relations have slightly larger slopes, so the effect of using these would be to lower the resulting gas masses. We note that the gas masses rely on the extinction correction based on the galaxy-integrated SED fitting and not on the \hahb measurements (which are uncertain for clump B). As mentioned in section 3.1, applying the Balmer decrement-based extinction correction would increase the gas masses by 50\%, the opposite effect of using the other KS laws. 

The statements made above rely on the KS conversion for normal Òmain sequenceÓ star-forming galaxies. While ZC406690 is in fact on the star-forming main sequence, there is some indication that the clumps might follow the ULIRG KS law, both because the clumps have similar properties to local starburst regions and also based on recent work by \cite{Sar+12} suggesting that \ztwo normal SFGs can simultaneously have a normal star-forming component and a starburst component. Using the ULIRG law to determine the gas masses for our clumps would lower the resulting masses. From \cite{Gen+10}, Figure 4, we find that the offset from the main-sequence KS law to the ULIRG law is around 0.8 dex for $\Sigma_{SFR}$s relevant for the clumps. Therefore, we assume a lower limit for the gas masses of 0.8 dex.

Fourth, we believe it is reasonable to assume the global KS relation for regions of $\sim$1 kpc in size. \cite{Bol+08}, find that many of the gas properties of extra-galactic GMCs are consistent with properties of the MW. In addition, \cite{Big+11} find that their binned 1 kpc-scale relation is almost identical to the relation binned from global literature data. \cite{Ken+07} also look at the KS relation on the scale of 0.5 and 2 kpc in M51a and find a very similar slope to that from the global relation from \cite{Ken98}, and an offset in zero-point of around 0.4 dex. It should be noted that the M51a integrated point falls 0.24 dex below the global law, so a good estimate for the offset in zero-point from extrapolation of the global to local KS relation would be around 0.15 dex. In addition, recent observations of the lensed cosmic eyelash galaxy at z$\sim$2.3 by \cite{Swi+11} and \cite{Dan+11}, probe the KS relation on the scale of 100 pc submillimeter-bright clumps. They find that the global KS relation should hold on scales of roughly $\sim$kpc, but should break down below 100 pc, and that globally, the galaxy is consistent with local ULIRGs (as expected for a high-z SMG). Finally, \cite{Ler+12} study H$\alpha$--, UV--, and IR-derived SFRs in addition to IRAM CO data at 1 kpc resolution for 30 disk galaxies, and find that the H$\alpha$--derived $\Sigma_{SFR}s$ could be off by about 0.3 dex at small vs. large scales, comparable to the 0.3 dex systematic uncertainty we assume.

The systematic uncertainties from applying the local KS relation to high-z galaxies, from the scatter between relations, from extrapolating to $\sim$kpc scales and from propagating the $\Sigma_{SFR}$ error are all less than or equal to the intrinsic scatter from the \cite{Big+11} and \cite{Gen+10} relations of 0.3 dex, so we therefore use this as the upper limit of the systematic uncertainty for our derived clump \siggas$~$ and as mentioned above, we use 0.8 dex as a lower limit. 

\section{Appendix B: Derivation of mass outflow rates}
In this section we discuss the model used to derive the mass outflow rates of the clumps, which is identical to the derivation used in \cite{Gen+11}. We derive the mass in the outflow and the mass outflow rates assuming two different geometric models. For both, we assume an outflow of warm, ionized gas into a solid angle $\Omega$ with radially constant outflow rate and velocity. As we are only calculating the mass outflow rate for the warm-ionized component and neglecting other components of the wind (neutral, molecular, and hot), our masses and rates are therefore lower limits. We assume case B recombination of photoionized gas with T = 10$^{4}$K, giving a volume emissivity of $\gamma_{H\alpha} = 3.56 \times 10^{-25}$ erg cm$^{-3}$ s$^{-1}$. We then calculate the mass of outflowing material from the extinction-corrected H$\alpha$ luminosity:

\begin{equation}
L_{H\alpha} = \gamma_{H\alpha} \int \! \Omega R^{2} n_{e}(R) n_{p}(R) \, \mathrm{d} R , \\
\end{equation}
\begin{equation}
M_{HII,He} = 1.36 \times m_{p} \int \! \Omega R^{2} n_{p} \, \mathrm{d} R = \frac{1.36 \times m_{p} L_{H\alpha}}{\gamma_{H\alpha} n_{eff}} , \\
\end{equation}
\begin{equation}
\dot{\rm M}_{\rm out} = \Omega R^{2} 1.36 \times m_{p} n(R) v_{out} = M_{HII,He} \times \frac{v_{out}}{R_{out}} \\
\end{equation}

where L$_{H\alpha}$ is the extinction-corrected \ha luminosity, v$_{out}$ is the outflow velocity derived from the \ha line profiles and given in Table 3, and R$_{out}$ is the maximum radial extent of the outflow, which is model dependent, as is n$_{eff}$ (electron density). The extinction-corrected \ha luminosity is taken from the broad-component of the line only, and is multiplied by an additional factor of 2 to account for the asymmetry of the broad line wing (i.e. much of the red wing is obscured, see Figure 2).

For our first model, we assume that the average electron density in the outflow scales like R$^{-2}$, but the local electron density of clouds and dense filaments in the outflow (from which we observe \ha emission) is constant at n$_{e}$ = 100 cm$^{-3}$.  In this model, R$_{out} \sim R_{HWHM}$ and is given in Table 2. As mentioned in section 4.1.4, the density in the wind could be closer to 4 cm$^{-3}$ and this is reflected in the uncertainties for \mdotout. In fact, the value of n$_{eff}$ is the dominant source of uncertainty for M$_{w}$ and \mdotout$~$ and for subsequent calculations that depend on these quantities. We find that the lower limit for n$_{eff}$ (10 cm$^{-3}$) gives unrealistic upper limits for the mass-loading factor of 30--60 (see Table 3).

For the second model, we assume that the entire outflow cone is filled with the ionized, H$\alpha$-emitting gas, and that n$_{eff}$ still scales like R$^{-2}$. Thus, n$_{eff} = n_{in}R_{in}/R_{out}$, where R$_{in}$ is the radius at which the outflow is launched and n$_{in}$ is the corresponding electron density at that radius. Assuming the outflow is very extended \citep{Ste+10}, we take R$_{out} \sim R_{disk} \sim 10 R_{in}$. From the n$_{e}^{2}$ scaling of the \ha emission, R$_{HWHM} \sim 2.3 R_{in}$. For n$_{in}$ we take an approximate average of the clump electron densities derived from the \sii$~$ doublet, $\sim 1000 ~ cm^{-3}$. Thus, for the second model, R$_{out}$ = 4.3 kpc and n$_{eff}$ = 100 cm$^{-3}$. These two models likely span the lower and upper limits for R$_{out}$ and so we take the average of the results from the two models as our wind mass and mass outflow rates (see Table 3).

\end{document}